\ifx\soda\undefined
    \documentclass[twoside,leqno]{article}
    \newcommand{\noSODA}[1]{#1}
    \newcommand{\inSODA}[1]{\commentout{#1}}
    \newcommand{\swapSODA}[2]{\commentout{#1}#2}
\else
    \documentclass[twoside,leqno,twocolumn]{article}
    \newcommand{\noSODA}[1]{\commentout{#1}}
    \newcommand{\inSODA}[1]{#1}
    \newcommand{\swapSODA}[2]{\commentout{#2}#1}
\fi

\usepackage{comment,amsfonts,amssymb,amsmath,amsthm,fullpage}
\newcommand{\commentout}[1]{}

\ifx\pdftexversion\undefined
\usepackage[colorlinks,linkcolor=black,filecolor=black,citecolor=black,urlcolor=black,pdfstartview=FitH]{hyperref}
\else
\usepackage[colorlinks,linkcolor=blue,filecolor=blue,citecolor=blue,urlcolor=blue,pdfstartview=FitH]{hyperref}
\fi

\newcommand{\alert}[1]{\textbf{\color{red}
[[[#1]]]}\marginpar{\textbf{\color{red}**}}\typeout{ALERT:
\the\inputlineno: #1}}

\newcommand{\dist}{{\rm dist}}

\newcommand{\bS}{{\bar{S}}}

\newcommand{\diam}{{\rm diam}}
\newcommand{\rad}{{\rm rad}}

\newcommand{\ie}{\emph{i.e. }}

\newcommand{\E}{{\mathbb{E}}}

\newcommand{\mommit}[1]{}
\newcommand{\namedref}[2]{\hyperref[#2]{#1~\ref*{#2}}}

\newcommand{\theoremref}[1]{\namedref{Theorem}{#1}}

\newcommand{\figureref}[1]{\namedref{Figure}{#1}}
\newcommand{\claimref}[1]{\namedref{Claim}{#1}}
\newcommand{\lemmaref}[1]{\namedref{Lemma}{#1}}

\newcommand{\eqnref}[1]{\namedref{Equation}{#1}}

\newtheorem{theorem}{Theorem}
\newtheorem{lemma}{Lemma}

\newtheorem{claim}[lemma]{Claim}
\newtheorem{definition}{Definition}

\begin{document}
\title{Embedding Metrics into Ultrametrics and Graphs into Spanning Trees with
Constant Average Distortion}

\author{Ittai Abraham\thanks{
        email: ittaia@cs.huji.ac.il.
    }\and Yair Bartal\thanks{
        email: yair@cs.huji.ac.il.
    Supported in part by a grant from the Israeli Science Foundation
(195/02).
    }\\
    Hebrew University \and Ofer Neiman\thanks{
        email: neiman@cs.huji.ac.il.
    Supported in part by a grant from the Israeli Science Foundation
(195/02).
    }}

\maketitle

\begin{abstract}
This paper addresses the basic question of how well can a tree
approximate distances of a metric space or a graph. Given a graph,
the problem of constructing a spanning tree in a graph which
strongly preserves distances in the graph is a fundamental problem
in network design. We present scaling distortion embeddings where
the distortion scales as a function of $\epsilon$, with the
guarantee that for each $\epsilon$ the distortion of a fraction
$1-\epsilon$ of all pairs is bounded accordingly. Such a bound
implies, in particular, that the \emph{average distortion} and
$\ell_q$-distortions are small. Specifically, our embeddings have
\emph{constant} average distortion and $O(\sqrt{\log n})$
$\ell_2$-distortion. This follows from the following results: we
prove that any metric space embeds into an ultrametric with scaling
distortion $O(\sqrt{1/\epsilon})$. For the graph setting we prove
that any weighted graph contains a spanning tree with scaling
distortion $O(\sqrt{1/\epsilon})$.  These bounds are tight even for
embedding in arbitrary trees.
 For probabilistic embedding into spanning trees
we prove a scaling distortion of $\tilde{O}(\log^2 (1/\epsilon))$,
which implies \emph{constant} $\ell_q$-distortion for every fixed
$q<\infty$.
\end{abstract}

\section{Introduction}\label{introduction}
  The problem of embedding general metric spaces into tree metrics
with small distortion has been central to the modern theory of
finite metric spaces. Such embeddings provide an efficient
representation of the complex metric structure by a very simple
metric. Moreover, the special class of ultrametrics (rooted trees
with equal distances to the leaves) plays a special role in such
embeddings \cite{bartal1,BLMN03}. Such an embedding provides an even
more structured representation of the space which has a hierarchical
structure \cite{bartal1}. Probabilistic embedding into ultrametrics
have led to algorithmic application for a wide range of problems
(see \cite{indyk}). An important problem in network design is to
find a tree spanning the network, represented by a graph, which
provides good approximation of the metric defined with the shortest
path distances in the graph. Different notions have been suggested
to quantify how well distances are preserved, e.g. routing trees and
communication trees \cite{bio}. The papers \cite{AKPW95,EEST05}
study the problem of constructing a spanning tree with low average
stretch, i.e., low average distortion over the edges of the tree. It
is natural to define our measure of quality for the embedding to be
its average distortion over all pairs, or alternatively the more
strict measure of its $\ell_2$-distortion. Such notions are very
common in most practical studies of embeddings (see for example
\cite{HS00,HF00-cofe,AS03,HBKKW03,ST04,TC04}) . We recall the
definitions from \cite{ABN06}:
 Given two metric spaces $(X,d_X)$ and $(Y,d_Y)$ an
\emph{injective} mapping $f:X \rightarrow Y$ is called an \emph{
embedding} of $X$ into $Y$. An embedding is \emph{non-contractive}
if for any $u \neq v \in X$: $d_Y(f(u),f(v)) \geq d_X(u,v)$. For a
non-contractive embedding let the distortion of the pair $\{u,v\}$
be $\dist_f(u,v) = \frac{d_Y(f(u),f(v))}{d_X(u,v)}$.
\begin{definition}[$\ell_q$-distortion]
For $1 \leq q \leq \infty$, define the \emph{$\ell_q$-distortion} of
an embedding $f$ as:
$$
\dist_q(f) = \| \dist_f(u,v) \|_q^{(\mathcal{U})} = \E[
\dist_f(u,v)^q ]^{1/q} ,
$$
where the expectation is taken according
to the uniform distribution ${\cal U}$ over ${X\choose 2}$. The
classic notion of \emph{distortion} is expressed by the
$\ell_\infty$-distortion and the \emph{average distortion} is
expressed by the $\ell_1$-distortion.
\end{definition}

The notion of average distortion is tightly related (see
\cite{ABN06}) to that of embedding with scaling distortion
\cite{KSW04,focs05,ABN06}.

\begin{definition}[Partial/Scaling Embedding]\label{def:partial-scaling}
Given two metric spaces $(X,d_X)$ and $(Y,d_Y)$, a \emph{partial
embedding} is a pair $(f,G)$, where $f$ is a non-contractive
embedding of $X$ into $Y$, and $G \subseteq {X\choose 2}$. The
distortion of $(f,G)$ is defined as: $\dist(f,G) = \sup_{\{u ,v\}
\in G} \dist_f(u,v)$.
For $\epsilon \in [0,1)$, a \emph{$(1-\epsilon)$-partial
embedding} is a partial embedding such that
$|G|\ge(1-\epsilon){n\choose 2}$.\footnote{Note that the embedding
is \emph{strictly} partial only if $\epsilon \geq 1/{n\choose 2}$.
}
Given two metric spaces $(X,d_X)$ and $(Y,d_Y)$ and a function
$\alpha: [0,1) \to \mathbb{R}^{+}$, we say that an embedding $f:X
\rightarrow Y$ has \emph{scaling distortion} $\alpha$ if for any $
\epsilon \in [0,1)$, there is some set $G(\epsilon)$ such that
$(f,G(\epsilon))$ is a $(1-\epsilon)$-partial embedding with
distortion at most $\alpha(\epsilon)$.
\end{definition}
We prove the following theorems:
%\newcounter{ThmScalingUltrametrics}
%\setcounter{ThmScalingUltrametrics}{\thetheorem}
\def\ThmScalingUltrametrics{
  Any $n$-point metric space embeds into an ultrametric with
  scaling distortion $O(\sqrt{1/\epsilon})$. In particular, its
  $\ell_q$-distortion is $O(1)$ for $1 \leq q <2$,
  $O(\sqrt{\log n})$ for $q=2$, and $O(n^{1-2/q})$ for $2< q \leq \infty$.
}
\begin{theorem}\label{thm:scaling-ultrametrics}
\ThmScalingUltrametrics
\end{theorem}
%\newcounter{ThmScalingSpanningTrees}
%\setcounter{ThmScalingSpanningTrees}{\thetheorem}
\def\ThmScalingSpanningTrees{
  Any weighted graph of size $n$ contains a spanning tree with
  scaling distortion $O(\sqrt{1/\epsilon})$. In particular, its
  $\ell_q$-distortion is $O(1)$ for $1 \leq q <2$,
  $O(\sqrt{\log n})$ for $q=2$, and $O(n^{1-2/q})$ for $2< q \leq \infty$.
}
\begin{theorem}\label{thm:scaling-spanning-trees}
\ThmScalingSpanningTrees
\end{theorem}
We show that the bounds in Theorems~\ref{thm:scaling-ultrametrics}
and~\ref{thm:scaling-spanning-trees} are tight for the $n$-node
cycle even for embeddings into arbitrary tree metrics. We also
obtain an equivalent result for probabilistic embedding into
spanning trees:
%\newcounter{ThmScalingProbSpanningTrees}
%\setcounter{ThmScalingProbSpanningTrees}{\thetheorem}
\def\ThmScalingProbSpanningTrees{
  Any weighted graph of size $n$ probabilistically embeds into a spanning tree with
  scaling distortion $\tilde{O}(\log^2{1/\epsilon})$. In particular, its
  $\ell_q$-distortion is $O(1)$ for any fixed $1 \leq q <
  \infty$.
}
\begin{theorem}\label{thm:scaling-prob-spanning-trees}
  Any weighted graph of size $n$ probabilistically embeds into a spanning tree with
  scaling distortion $\tilde{O}(\log^2{1/\epsilon})$. In particular, its
  $\ell_q$-distortion is $O(1)$ for any fixed $1 \leq q <
  \infty$\footnote{Note that probabilistic embedding bounds on the $\ell_q$-distortion \emph{do not}
imply an embedding into a single tree with the same bounds, with the
exception of $q=1$.}.

\end{theorem}
\subsection{Related Work}
Embedding metrics into trees and ultrametrics was introduced in the
context of probabilistic embedding in \cite{bartal1}. Other related
results on embedding into ultrametrics include work on metric Ramsey
theory \cite{BLMN03}, multi-embeddings \cite{bm} and dimension
reduction \cite{bm:ultradim}. Embedding an arbitrary metric into a
tree metric requires $\Omega(n)$ distortion in the worst case even
for the metric of the $n$-cycle \cite{RR98}. It is a simple fact
\cite{HM06,BLMN03,bartal1} that any $n$-point metric embeds in an
ultrametric with distortion $n-1$. However the known constructions
are not scaling and have average distortion linear in $n$. The
probabilistic embedding theorem \cite{FRT,Bartal04} (improving
earlier results of \cite{bartal1,bartal2}) states that any $n$-point
metric space probabilistically embeds into an ultrametric with
distortion $O(\log n)$. This result has been the basis to many
algorithmic applications (see \cite{indyk}). This theorem implies
the existence of a single ultrametric with average distortion
$O(\log n)$ (a constructive version was given in \cite{Bartal04}).
This bound was later improved with the analysis of \cite{focs05} as
we discuss below. The study of partial embedding and scaling
distortion was initiated by Kleinberg, Slivkins and
Wexler~\cite{KSW04}, and later studied in \cite{focs05,ABN06}.
Abraham et. al \cite{focs05} prove that any finite metric space
probabilistically embeds in an ultrametric with scaling distortion
$O(\log(1/\epsilon))$ implying constant average distortion. As
mentioned above, since the distortion is bounded in expectation,
this result implies the existence of a single ultrametric with
constant average distortion, but does not bound the
$\ell_2$-distortion. In \cite{ABN06} we have studied in depth the
notions of average distortion and $\ell_q$-distortion and their
relation to partial and scaling embeddings. Our main focus was the
study of optimal scaling embeddings for embedding into $L_p$ spaces.
For embedding of metrics into ultrametrics, we mentioned that
\emph{partial} embeddings exist with distortion
$O(\sqrt{1/\epsilon})$ matching the lower bound from \cite{focs05}.
\theoremref{thm:scaling-ultrametrics} significantly strengthens this
result by providing an embedding with \emph{scaling} distortion.
That is, the bound holds for all values of $0< \epsilon < 1$
\emph{simultaneously} and therefore the embedding has bounded
$\ell_q$-distortion.
 It is a basic fact
that the minimum spanning tree in an $n$-point weighted graph
preserves the (shortest paths) metric associated with the graph up
to a factor of $n-1$ at most. This bound is tight for the
$n$-cycle. Here too, it is easy to see that the MST does not have
scaling distortion, and may result in linear average distortion.
Alon, Karp, Peleg and West \cite{AKPW95} studied the problem of
computing a spanning tree of a graph with small average stretch
(over the edges of the graph). This can also be viewed as the dual
of probabilistic embedding of the graph metric in spanning trees.
Their work was recently significantly improved by Elkin, Emek,
Speilman and Teng \cite{EEST05} who show that any weighted graph
contains a spanning tree with average stretch $O(\log^2 n\log\log
n)$. This result can also be rephrased in terms of the average
distortion (but not the $\ell_2$-distortion) over all pairs.
 For spanning trees, this paper gives the
\emph{first} construction with \emph{constant} average distortion.

\subsection{Discussion of Techniques}
\theoremref{thm:scaling-ultrametrics} uses partitioning techniques
similar  to those used in the context of the metric Ramsey problem
\cite{bbm,BLMN03}. However, in our case we need to provide an
argument for the existence of a partition which simultaneously
satisfies multiple conditions, each for every possible value of
$\epsilon$. \theoremref{thm:scaling-spanning-trees} builds on the
technique above together with the Elkin et. al. \cite{EEST05} method
to construct a spanning tree. A straightforward application of this
approach loses an extra $O(\log n)$ factor and hence does not give a
scaling distortion depending solely on $\epsilon$. The loss in the
Elkin et.al. approach stems from the need to bound the diameter in
the recursive construction of the spanning tree. In each level of
the construction we may alow only a very small increase as these get
multiplied in the bound on the total blow up in the overall
diameter. In their original work \cite{EEST05} the increase per
level is $\Theta(1/\log n)$ which translates to the blow up in the
distortion. In our case we show that the increase can exponentially
decrease along the levels. This indeed guarantees a good blow up in
the overall diameter but is awful in terms of the distortion. We
apply a new technique for bounding the diameter which allows us to
limit the number of levels involved. On the other hand it is clear
that for every value of $\epsilon$ there is a limited number of
levels for which the distortion requirement imposes new constraints.
The proof then proceeds to carefully balance these different
arguments. \theoremref{thm:scaling-prob-spanning-trees} uses
essentially the same ideas together with the known probabilistic
embedding methods (in fact, the proof of this theorem is somewhat
less technically involved). \inSODA{In this extended abstract we
focus mainly on the proofs on
Theorems~\ref{thm:scaling-ultrametrics}
and~\ref{thm:scaling-spanning-trees}.

%and only sketch the proof of
%\theoremref{thm:scaling-prob-spanning-trees}.

The Full version of the proofs can be found in \cite{ABN06-TR}.} The
fact that these theorems are tight essentially follows from the
results and techniques of \cite{focs05,ABN06}.

\section{Preliminaries}
Consider a finite metric space $(X,d)$ and let $n=|X|$. For any
point $x\in X$ and a subset $S\subseteq X$ let $d(x,S)=\min_{s\in
S}d(x,s)$. The \emph{diameter} of $X$ is denoted
$\diam(X)=\max_{x,y\in X} d(x,y)$. For a point $x\in X$ and $r \ge
0$, the ball at radius $r$ around $x$ is defined as $B_X(x,r) =
\{z \in X | d(x,z) \leq r \}$. We omit the subscript $X$ when it
is clear form the context. Given $x \in X$ let $\rad_x(X)=\max_{y
\in X} d(x,y)$. When a cluster $X$ has a center $x \in X$ that is
clear from the context we will omit the subscript and write
$\rad(X)$ instead of $\rad_x(X)$. Given an edge-weighted graph
$G=(X,E,\omega)$ with $\omega:E \to \mathbb{R}^+$, let $(X,d)$ be
the  metric space induced from the graph in the usual manner -
vertices are associated with points, distances between points
correspond to shortest-path distances in $G$.
\begin{definition}\label{def:UM}
An ultrametric $U$ is a metric space $(U,d_U)$ whose elements are
the leaves of a rooted labelled tree $T$. Each $v\in T$ is
associated a label $\Phi(v) \geq 0$ such that if $u \in T$ is a
descendant of $v$ then $\Phi(u) \le \Phi(v)$ and $\Phi(u)=0$ iff
$u \in U$ is a leaf. The distance between leaves $x,y \in U$ is
defined as $d_U(x,y)=\Phi({\rm lca}(x,y))$ where ${\rm lca}(x,y)$
is the least common ancestor of $x$ and $y$ in $T$.
\end{definition}

\section{Scaling embedding into an ultrametric}
\begin{theorem}
\ThmScalingUltrametrics
\end{theorem}
We give the proof for scaling distortion. The consequence of the
bounds on the $\ell_q$-distortion follows by a simple calculation.
The proof is by induction on the size of $X$ (the base case is where
$|X|=1$ and is trivial). Assume the claim is true for any metric
space with less than $n$ points. Let $(X,d)$ be a metric space with
$n=|X|$ and $\Delta = \diam(X)$. The ultrametric is defined in a
standard manner by defining the labelled tree $T$ whose leaf-set is
$X$. The high level construction of $T$ is as follows: find a
partition $P$ of $X$ into $X_1$ and $X_2 = X \setminus X_1$, the
root of $T$ will be labelled $\Delta$, and its children $T_1,T_2$
will be the trees formed recursively from the ultrametric trees of
$X_1$ and $X_2$ respectively. Let $u\in X$ be such that
$|B(u,\Delta/2)|\le n/2$ (such a point can always be found). For any
$0<\epsilon\le 1$ denote by $B_\epsilon(X)$ the total number of
pairs $(x,y)\in X$ such that $d_T(x,y)
> (150 / \sqrt{\epsilon}) d_X(x,y)$.
For a partition $P=(X_1;X_2)$ let
$\hat{B}_\epsilon(P)=|\{(x,y)\mid x\in X_1\wedge y\in X_2\wedge
d_X(x,y) \le (\sqrt{\epsilon}/150)\cdot\Delta\}|$.
\begin{claim}\label{claim:bad-num}
Let $\epsilon \in (0,1]$ and let $(X,d)$ be a metric space, if for
any sub metric $X' \subseteq X$ there exists a partition
$P=(X_1;X_2)$ be a partition of $X'$ such that $\hat{B}_\epsilon(P)<
\epsilon|X_1|\cdot|X_2|$ then $B_\epsilon(X)\le\epsilon{|X|\choose
2}$.
\end{claim}
\begin{proof}
Let $P=(X_1;X_2)$ be a partition of $X$ such that
$\hat{B}_\epsilon(P)\le\epsilon|X_1|\cdot|X_2|$. By induction,
\begin{eqnarray*}
B_\epsilon(X)  &\leq& \hat{B}_\epsilon(P)+ B_\epsilon(X_1) +
B_\epsilon(X_2)\\
&\leq&  \epsilon\left({|X_1|\choose 2}+{|X_2|\choose 2} +
|X_1|\cdot|X_2|\right)\\
\noSODA{&=&\epsilon/2\left(|X_1|^2-|X_1|+|X_2|^2-|X_2|+2|X_1|\cdot|X_2|\right)\\}
&=& \epsilon/2\left((|X_1|+|X_2|)(|X_1|+|X_2|-1)\right)\\
&=& \epsilon {|X|\choose 2}.
\end{eqnarray*}
\end{proof}
So it is sufficient to show that there exists a partition satisfying
\claimref{claim:bad-num} for all $\epsilon \in (0,1]$
simultaneously.
\paragraph{Partition Algorithm.}
Let $\hat{\epsilon} = \max \{\epsilon \in (0,1] \mid
|B(u,\sqrt{\epsilon}\Delta/4)| \ge \epsilon n\}$. Observe that
$1/n\le\hat{\epsilon}\le 1/2$ by the choice of $u$ . Define the
intervals
$\hat{S}=[\sqrt{\hat{\epsilon}}\Delta/4,\sqrt{\hat{\epsilon}}\Delta/2]$,
$S=[(\frac{1}{4}+\frac{1}{25})\sqrt{\hat{\epsilon}}\Delta,(\frac{1}{2}-\frac{1}{25})\sqrt{\hat{\epsilon}}\Delta]$,
$s=\frac{17}{100}\sqrt{\hat{\epsilon}}\Delta$, and the shell
$Q=\{w\mid d(u,w)\in \hat{S}\}$. We partition $X$ by choosing some
$r \in S$ such that $X_1=B(u,r)$ and $X_2=X\setminus X_1$. The
following property will be used in several cases:
\begin{claim} \label{claim:small-x1}
$|B(u,\sqrt{\hat{\epsilon}}\Delta/2)| \leq 4\hat{\epsilon}n$.
\end{claim}
\begin{proof}
There are two cases: If $\hat{\epsilon} \leq  1/4$ then
$|B(u,\sqrt{\hat{\epsilon}}\Delta/2)|=
|B(u,\sqrt{4\hat{\epsilon}}\Delta/4)|\leq 4\hat{\epsilon}n$
(otherwise contradiction to maximality of $\hat{\epsilon}$).
Otherwise, $\hat{\epsilon}  \in (1/4, 1]$. In such a case
$|B(u,\sqrt{\hat{\epsilon}}\Delta/2)| \le |B(u,\Delta/2)| \le n/2\le
2\hat{\epsilon}n$.
\end{proof}
We will now show that some choice of $r\in S$ will produce a
partition that satisfies \claimref{claim:bad-num} for all $\epsilon
\in  (0, 32\hat{\epsilon}]$. For any $r \in S$ and $\epsilon \leq
32\hat{\epsilon}$ let $S_r(\epsilon) =
(r-\sqrt{\epsilon}\Delta/150,r+\sqrt{\epsilon}\Delta/150)$,
$s(\epsilon)=\sqrt{\epsilon}\Delta/75$, and let $Q_r(\epsilon)=\{w
\mid d(u,w)\in S_r(\epsilon)\}$. Notice that for any $r\in S$ and
any $\epsilon\le 32\hat{\epsilon}$ : $S_r(\epsilon)\subseteq
\hat{S}$. Define that properly $A_r(\epsilon)$ holds if cutting at
radius $r$ is ``good" for $\epsilon$, formally: $A_r(\epsilon)$ iff
$|Q_r(\epsilon)| < \sqrt{\epsilon\cdot\hat{\epsilon}/2}\cdot n$. For
any $\epsilon \leq 32\hat{\epsilon}$, note that in any partition to
$X_1=B(u,r)$, $X_2=X\setminus X_1$ only pairs $(x,y)$ such that
$x,y\in Q_r(\epsilon)$ are distorted by more than
$O(\sqrt{1/\epsilon})$. If property $A_r(\epsilon)$ holds then
$\hat{B}_\epsilon(P) \le \epsilon \cdot \hat{\epsilon}n^2/2$. Since
$\hat{\epsilon} n \leq  |X_1| \leq n/2$ then $\epsilon \cdot
\hat{\epsilon}n^2/2 \leq \epsilon n/2 |X_1| \leq \epsilon |X_1|
|X_2|$ so $A_r(\epsilon)$ implies \claimref{claim:bad-num} for
$\epsilon$. Hence for $\epsilon \in (0, 32\hat{\epsilon}]$ the
following is sufficient:
\begin{claim}\label{claim:small-eps}
There exists some $r\in S$ such that properly $A_r(\epsilon)$
holds for all $\epsilon \in (0, 32\hat{\epsilon}]$.
\end{claim}
\begin{proof}
The proof is based on the following iterative process that
greedily deletes the ``worst" interval in $S$. Initially, let
$I_0=S$, and $j=1$:
\begin{enumerate}

\item If for all $r \in I_{j-1}$ and for all $\epsilon \le
32\hat{\epsilon}$ property $A_r(\epsilon)$ holds then set $t=j-1$,
stop the iterative process and output $I_t$.

\item Let $\mathcal{S}_j=\{S_r(\epsilon) \mid r \in I_{j-1},
\epsilon \leq 32 \hat{\epsilon}, \neg A_r(\epsilon)\}$. We greedily
remove the interval $S \in \mathcal{S}_j$ that has maximal
$\epsilon$. Formally, let $r_j,\epsilon_j$ be parameters such that
$S_{r_j}(\epsilon_j) \in \mathcal{S}_j$ and $\epsilon_j = \max \{
\epsilon \mid \exists S_r(\epsilon) \in \mathcal{S}_j \}$.

\item Set $I_j = I_{j-1}\setminus S_{r_j}(\epsilon_j)$, set
$j=j+1$, and goto 1.
\end{enumerate}

Let $ \mathcal{Q} = \{ Q_r(\epsilon) \}$ and note that
$|\mathcal{Q}|=O(n^2)$ and it is easy to show that for every $j \in
\{1,\dots,t\}$, $Q' \in \mathcal{Q}$, the maximum of $\{\epsilon
\mid S_r(\epsilon) \in \mathcal{S}_j, Q_r(\epsilon)=Q\}$ is obtained
inside the set and can be found in $O(n^2)$ time.

We now argue that $I_t\neq\emptyset$ and hence such a value $r \in
S$ can be found. Since for any $1\le j < i \le t$, $s(\epsilon_j)
\ge s(\epsilon_i)$ it follows that any $x \in Q$ appears in at most
2 ``bad" intervals. From this and \claimref{claim:small-x1}:
\[
\sum_{j=1}^t|Q_{r_j}(\epsilon_j)| \le 2|Q|\le 8\hat{\epsilon}n.
\]
Recall that since $A_{r_j}(\epsilon_j)$ does not hold then for any
$1\le j\le t$ : $|Q_{r_j}(\epsilon_j)| \ge
\sqrt{\epsilon_j\cdot\hat{\epsilon}/2}\cdot n$ which implies that
\[
\sum_{j=1}^t\sqrt{\epsilon_j} \le 12\sqrt{\hat{\epsilon}}.
\]
On the other hand, by definition
\[
\sum_{j=1}^t s(\epsilon_j)\le
\sum_{j=1}^t\sqrt{\epsilon_j}\Delta/75\le
12/75\cdot\sqrt{\hat{\epsilon}}\Delta =
16/100\cdot\sqrt{\hat{\epsilon}}\Delta.
\]
Since $s=17/100\cdot \sqrt{\hat{\epsilon}}\Delta$ then indeed
$I_t\neq\emptyset$ so any $r\in I_t$ satisfies the condition of
the claim.
\end{proof}
It remains to show that any choice of $r\in S$ will produce a
partition that satisfies \claimref{claim:bad-num} for all $\epsilon
\in  (32\hat{\epsilon},1]$.
\begin{claim}\label{claim:big-epsilon}
If $\epsilon \in (32\hat{\epsilon},1]$, $r \in S$ and $P=(B(u,r);
X \setminus B(u,r))$ then $\hat{B}_\epsilon(P) < \epsilon
|X_1|\cdot|X_2|$.
\end{claim}
\begin{proof}
Let $\epsilon \in (32\hat{\epsilon},1]$ and fix some $r\in S$.
Only pairs $(x,y)$ such that $x\in X_1$ and $y\in
B(u,r+\sqrt{\epsilon}\Delta/16)\cap X_2$ can be distorted by more
than $16\sqrt{1/\epsilon}$ and hence may be counted in
$\hat{B}_\epsilon(P)$.
Since $\sqrt{\hat{\epsilon}}\le\sqrt{\epsilon/2}/4$ and
$r<\sqrt{\hat{\epsilon}}\Delta/2$ then
$|B(u,r+\sqrt{\epsilon}\Delta/16)| \le
|B(u,\sqrt{\epsilon/2}(\frac{1}{8}+\frac{1}{8})\Delta)| =
|B(u,\sqrt{\epsilon/2}\Delta/4)| < \epsilon n/2$ by the maximality
of $\hat{\epsilon}$.
Since $|X_2|\ge n/2$ it follows that $\hat{B}_\epsilon(P)\leq
\epsilon |X_1|\cdot|X_2|$, as required.
\end{proof}
\begin{proof}[Proof of \theoremref{thm:scaling-ultrametrics}]
{}From \claimref{claim:small-eps} and \claimref{claim:big-epsilon},
it follows that our partition scheme finds a cut $P=(X_1; X_2)$ such
that $\hat{B}_\epsilon(P) < \epsilon |X_1|\cdot|X_2|$ for all
$\epsilon$. Hence when applying the partition scheme inductively, by
\claimref{claim:bad-num} the theorem follows.
\end{proof}

\section{Scaling Embedding into a Spanning Tree}
Here we extended the techniques of the previous section, in
conjunction with the constructions of \cite{EEST05} to achieve the
following:
\begin{theorem}
\ThmScalingSpanningTrees
\end{theorem}
Given a graph, the spanning tree is created by recursively
partitioning the metric space using a \emph{hierarchical star
partition}. The algorithm has three components, with the following
high level description:

1. A decomposition algorithm that creates a single cluster. The
decomposition algorithm is similar in spirit to the decomposition
algorithm used in the previous section for metric spaces. We will
later explain the main differences.

2. A star partition algorithm. This algorithm partitions a graph
$X$ into a central ball $X_0$ with center $x_0$ and a set of cones
$X_1,\dots,X_m$ and also outputs a set of edges of the graph
$(y_1,x_1),\dots,(y_m,x_m)$ that connect each cone set, $x_i \in
X_i$ to the central ball, $y_i \in X_0$. The central ball is
created by invoking the decomposition algorithm with a center $x$
to obtain a cluster whose radius is in the range
$[(1/2)\rad_{x_0}(X) \dots (5/8)\rad_{x_0}(X)]$. Each cone set
$X_i$ is created by invoking the decomposition algorithm on the
``cone-metric" obtained from $x_0,x_i$. Informally, a ball in the
cone-metric around $x_i$ with radius $r$ is the set of all points
$x$ such that $d(x_0,x_i)+d(x_i,x)-d(x_0,x)\leq r$. Hence each
cone $X_i$ is a ball whose center is $x_i$ in some appropriately
defined ``cone-metric". The radius of each ball in the cone metric
is chosen to be $\approx \tau^k \rad_{x_0}(X)$ where $\tau<1$ is
some fixed constant and $k$ is the depth of the recursion.
Unfortunately, at some stage the radius may be too small for the
decompose algorithm to preform well enough. In such cases we must
reset the parameters that govern the radius of the cones. (in the
next bullet, we will define more accurately how the recursion is
performed and when this parameter of a cluster may be reset). The
main property of this star decomposition is that for any point $x
\in X_i$, the distance to the center $x_0$ does not increase by
too much. More formally, $d_{X_0 \cup \{(y_i,x_i)\} \cup
X_i}(x_0,x)/d(x_0,x) \leq \prod_{j\leq k} (1+\tau^j)$ where $k$ is
the depth of the recursion. Informally, this property is used in
order to obtain a constant blowup in the diameter of each cluster
in the final spanning tree.

3. Recursive application of the star partition. As mentioned in
the previous bullet, the radius of the balls in the cone metric
are exponentially decreasing. However at certain stages in the
recursion, the cone radius becomes too small and the parameters
governing the cone radius must be reset. Clusters in which the
parameters need to be restarted are called \emph{reset clusters}.
The two parameters that are associated with a reset cluster $X$
are $n=|X|$, and $\Lambda = \rad(X)$. Specifically, a cluster is
called a reset cluster if its size relative to the size of the
last reset cluster is larger than some constant times its radius
relative to radius of the last reset cluster. In that case $n$ and
$\Lambda$ are updated to the values of the current cluster. This
implies that reset clusters have small diameter, hence their total
contribution to the increase of radius is small. Moreover,
resetting the parameters allows the decompose algorithm to
continue to produce the clusters with the necessary properties to
obtain the desired scaling distortion. Using resets, the algorithm
can continue recursively in this fashion until the spanning tree
is formed.

\paragraph{Decompose algorithm.}
The decompose algorithm receives as input several parameters.
First it obtains a pseudo-metric space $(W,d)$ and point $u$ (for
the central ball this is just the shortest-paths metric, while for
cones, this pseudo metric is the so called ``cone-metric" which
will be formally defined in the sequel). The goal of the decompose
algorithm is to partition $W$ into a cluster which is a ball
$Z=B(u,r)$ and $\bar{Z}=W \setminus Z$.

Informally, this partition $P$ is carefully chosen to maintain the
scaling property: for every $\epsilon$, the number of pairs whose
distortion is too large is ``small enough". Let $\hat{\Lambda}$ be a
parameter corresponding to the radius of the cluster over which the
star-partition is performed. Pairs that are separated by the
partition may risk the possibility of being at distance
$\Theta(\hat{\Lambda})$ in the constructed spanning tree. We denote
by $\hat{B}_\epsilon(P)$ the number of pairs that may be distorted
by at least $\Omega(\sqrt{1/\epsilon})$
%\alert{Ofer: I think it should be little omega..}
if the distance between them will grow to $\hat{\Lambda}$. There are
several parameters that control the number of pairs in
$\hat{B}_\epsilon(P)$. Given a parameter $n \geq |W|$ which
corresponds to the size of the last reset cluster containing $W$, we
expect the number of ``bad" pairs for a specific value of $\epsilon$
to be at most $O(\epsilon |Z| \cdot (n-|Z|))$. To allow to control
this bound even tighter we have an additional parameter $\beta$ so
that the partition $P$ will have the property that
$\hat{B}_\epsilon(P) = O(\epsilon |Z| \cdot (n-|Z|) \cdot \beta)$.
However, if we insist that this property holds true for all
$\epsilon$ we cannot maintain a small enough bound on the maximum
value for the radius $r$. Since this value determines the amount of
increase in the radius of the cluster, we would like to be able to
bound it. Therefore, we keep another parameter, denoted
$\epsilon_{\rm lim}$. That is, the partition $P$ will be good only
for those values of $\epsilon$ satisfying $\epsilon \leq
\epsilon_{\rm lim}$.

The radius $r$ of the ball is controlled by the parameters
$\hat{\Lambda}$, $\theta$ and a value $\alpha \leq
\sqrt{\epsilon_{\rm lim}}$. The guarantee is that $r \in
[\theta\hat{\Lambda},(\theta +\alpha)\hat{\Lambda}]$. Recall that
$\hat{\Lambda}$, corresponds to the radius of the cluster over
which the star-partition is performed. For the central ball of the
star-partition $\theta$ is fixed to $1/2$ and for the star's cones
$\theta$ is fixed to $0$. Indeed, as indicated above, the value of
$\epsilon_{\rm lim}$ determines the increase in the radius of the
cluster by setting the value for $\alpha$. This cannot, however,
be set arbitrarily small, in order to satisfy all of the
partition's properties, and so $\epsilon_{\rm lim}$ must be set
above some minimum value of $|W|/(n\cdot \beta)$. Intuitively, we
can only keep $\alpha$ small if $|W| \ll n$.

Let us explain now how the decompose algorithm will be used within
our overall scheme. The parameter $\beta$ is chosen such that it
is bounded by $\mu^{k}$ where $\mu <1$ is some fixed constant and
$k$ is the depth of the recursion from the last reset cluster.
Hence, for every $\epsilon$ that is smaller than $\epsilon_{\rm
lim}$, the property obtained by the decompose algorithm is that
the number of newly distorted edges is at most $O(\epsilon |Z|
\cdot (n-|Z|) \cdot \mu^k)$. For $\epsilon$ that are larger than
$\epsilon_{\rm lim}$, we show that the number of points in the
current cluster is less than an $\epsilon$ fraction of the number
of points in the last reset cluster, hence we can discard all the
pairs in such clusters and the total sum of all such discarded
pairs is small. Therefore, the total number of distorted edges is
bounded by summing the distorted edges over all clusters, for each
cluster depending on whether $\epsilon$ is smaller or larger than
$\epsilon_{\rm lim}$ of that cluster. The bound obtained also uses
the fact that $\mu^k$ is a geometric series.

Now, if $X$ is not a reset cluster then $|X|/n$ is small compared to
the ratio of its radius and the radius of the last reset cluster. We
show that this ratio drops exponentially, bounded by
$(\frac{5}{8})^k$, where $k$ is the depth of the recursion since the
last reset cluster. By letting $\epsilon_{\rm lim} = |X|/(n\cdot
\beta)$, and as $\mu < \frac{5}{8}$, we maintain that $\alpha  \leq
\sqrt{\epsilon_{\rm lim}} = \tau^k$ for some $\tau<1$, as we
desired.

We now turn to the formal description of the algorithm and its
analysis. We will make use of the following predefined constants:
$c=2e$, $c'=e(2e+1)$, $\hat{c} = 22$, and $C=8\sqrt{c\cdot
\hat{c}}$. Finally, the distortion is given by $\hat{C} = 150C\cdot
c'$. For any $0<\epsilon\le 1$ denote by $B_\epsilon(X)$ the total
number of pairs $(x,y)\in X$ such that $d_T(x,y) > ( \hat{C}/
\sqrt{\epsilon}) d_X(x,y)$. The exact properties of the
decomposition algorithm is captured by the following Lemma:

\begin{lemma}\label{lem:partition}
Given a metric space $(W,d)$, a point $u \in W$ and parameters $n
\in \mathbb{N}$, $\hat{\Lambda}>0$, and $\beta,\theta >0$, there
exists an algorithm $\texttt{decompose}((W,d),u,
\hat{\Lambda},\theta, n, \epsilon_{\rm lim},\beta)$ that computes
a partition $P=(Z;\bar{Z})$ of $W$ such that $Z=B_{(W,d)}(u,r)$
and $r/\hat{\Lambda}\in[\theta,\theta +\alpha]$ where $\alpha=
\sqrt{\epsilon_{\rm lim}}/C$. Let
$\hat{B}_\epsilon(P)=|\{(x,y)\mid x\in Z\wedge y\in\bar{Z}\wedge
d(x,y) \le\frac{\sqrt{\epsilon}\cdot\hat{\Lambda}}{150C} \}|$. For
$n \geq |W|$ and $\epsilon_{\rm lim} \geq \frac{|W|}{\beta \cdot
n}$ the partition has the property that for any
$\epsilon\in(0,\epsilon_{\rm lim}]$:
\[
\hat{B}_\epsilon(P)\le \epsilon|Z|\cdot(n-|Z|)\cdot\beta.
\]
\end{lemma}
\paragraph{Star-Partition algorithm.}
Consider a cluster $X$ with center $x_0$ and parameters $n,
\Lambda$. Recall that parameters $n,\Lambda$ are the number of
points and the radius (respectively) of the last reset cluster. A
star-partition, partitions $X$ into a central ball $X_0$, and
cone-sets $X_1,\dots,X_m$ and edges $(y_1,x_1),\dots,(y_m,x_m)$,
the value $m$ is determined by the star-partition algorithm when
no more cones are required. Each cone-set $X_i$ is connected to
$X_0$ by the edge $(y_i,x_i), y_i\in X_0, x_i\in X_i$. Denote by
$P_0$ the partition creating the central ball $X_0$ and by
$\{P_i\}_{i=1}^m$ the partitions creating the cones. In order to
create the cone-set $X_i$ use the decompose algorithm on the
cone-metric $\ell^{x_0}_{x_i}$ defined below.
\begin{definition}[cone metric\footnote{In fact, the cone-metric is a
pseudo-metric.}] Given a metric space $(X,d)$ set $Y \subset X$,
$x\in X$, $y \in Y$ define the \emph{cone-metric }$\ell^x_y:Y^2\to
\mathbb{R}^+$ as
$\ell^x_y(u,v)=|(d(x,u)-d(y,u))-(d(x,v)-d(y,v))|$.
\end{definition}
Note that $B_{(Y,\ell^x_y)}(y,r) = \{ v \in Y | d(x,y)+ d(y,v) -
d(x,v) \leq r \}$.

\swapSODA{
\begin{figure}[ht!]
\small{\fbox{
\begin{minipage}[t]{90mm}
$(X_0,\dots,X_m,(y_1,x_1),\dots,(y_m,x_m))=$\\$\texttt{star-partition}(X,x_0,n,\Lambda)$:
\begin{enumerate}
\item Set $i=0$ ; $\beta =
\frac{1}{\hat{c}}\left(\frac{\rad_{x_0}(X)}{\Lambda}\right)^{1/4}$;
$\epsilon_{\rm lim} = |X|/(\beta n)$; $\hat{\Lambda}=\rad_{x_0}(X)$;

\item $(X_i,Y_i) =
\texttt{decompose}((X,d),x_0,\hat{\Lambda},1/2,n,\epsilon_{\rm
lim},\beta)$;
\item If $Y_i = \emptyset$ set $m=i$ and stop; Otherwise, set
$i=i+1$;

\item Let $(x_i,y_i)$ be an edge in $E$ such that $y_i\in
X_0,x_i\in Y_{i-1}$;

\item Let $\ell=\ell^{x_0}_{x_i}$ be cone-metric of $x_0,x_i$ on the subspace $Y_{i-1}$;

\item $(X_i,Y_i) = \texttt{decompose}((Y_{i-1},
\ell),x_i,\hat{\Lambda},0,n,\epsilon_{\rm lim},\beta)$;

\item goto 3;
\end{enumerate}
\end{minipage}}}\caption{\texttt{star-partition} algorithm} \label{fig:star}
\end{figure}

\paragraph{Hierarchical-Star-Partition algorithm.}
Given a graph $G=(X,E,\omega)$, create the tree by choosing some
$x \in X $, setting $X$ as a reset cluster and calling:
$\texttt{hierarchical-star-partition}(X,x,|X|,\rad_x(X))$.
\begin{figure}[ht!]
\small{\fbox{
\begin{minipage}[t]{92mm}
$T=\texttt{hierarchical-star-partition}(X,x,n,\Lambda)$:
\begin{enumerate}
\item If $|X|=1$ set $T=X$ and stop.

\item
$(X_0,\dots,X_m,(y_1,x_1),\dots,(y_m,x_m)) =$\\$
\texttt{star-partition}(X,x,n,\Lambda)$;

\item For each $i \in [1,\dots,m]$:

\item If $\frac{|X_i|}{n}\le
c\frac{\rad_{x_i}(X_i)}{\Lambda}$ then \\ $T_i =
\texttt{hierarchical-star-partition}(X_i,x_i,n,\Lambda)$;

\item
Otherwise, set $X_i$ to be a \textbf{reset cluster}, \\
$T_i =
\texttt{hierarchical-star-partition}(X_i,x_i,|X_i|,\rad_{x_i}(X_i))$;

\item Let $T$ be the tree formed by connecting $T_0$ with $T_i$
using edge $(y_i,x_i)$ for each $i \in [1,\dots,m]$;
\end{enumerate}
\end{minipage}}}\caption{\texttt{hierarchical-star-partition} algorithm} \label{fig:hirachical-tar}
\end{figure}
}
{
\begin{figure}[ht!]
\fbox{
\begin{minipage}[t]{180mm}
$(X_0,\dots,X_m,(y_1,x_1),\dots,(y_m,x_m))=\texttt{star-partition}(X,x_0,n,\Lambda)$:
\begin{enumerate}
\item Set $i=0$ ; $\beta =
\frac{1}{\hat{c}}\left(\frac{\rad_{x_0}(X)}{\Lambda}\right)^{1/4}$;
$\epsilon_{\rm lim} = |X|/(\beta n)$; $\hat{\Lambda}=\rad_{x_0}(X)$;

\item $(X_i,Y_i) =
\texttt{decompose}((X,d),x_0,\hat{\Lambda},1/2,\epsilon_{\rm
lim},\beta)$;
\item If $Y_i = \emptyset$ set $m=i$ and stop; Otherwise, set
$i=i+1$;

\item Let $(x_i,y_i)$ be an edge in $E$ such that $y_i\in
X_0,x_i\in Y_{i-1}$;

\item Let $\ell=\ell^{x_0}_{x_i}$ be cone-metric of $x_0,x_i$ on the subspace $Y_{i-1}$;

\item $(X_i,Y_i) = \texttt{decompose}((Y_{i-1},
\ell),x_i,\hat{\Lambda},0,\epsilon_{\rm lim},\beta)$;

\item goto 3;
\end{enumerate}
\end{minipage}}\caption{\texttt{star-partition} algorithm} \label{fig:star}
\end{figure}

\paragraph{Hierarchical-Star-Partition algorithm.}
Given a graph $G=(X,E,\omega)$, create the tree by choosing some
$x \in X $, setting $X$ as a reset cluster and calling:
$\texttt{hierarchical-star-partition}(X,x,|X|,\rad_x(X))$.
\begin{figure}[ht!]
\fbox{
\begin{minipage}[t]{180mm}
$T=\texttt{hierarchical-star-partition}(X,x,n,\Lambda)$:
\begin{enumerate}
\item If $|X|=1$ set $T=X$ and stop.

\item
$(X_0,\dots,X_m,(y_1,x_1),\dots,(y_m,x_m)) =
\texttt{star-partition}(X,x,n,\Lambda)$;

\item For each $i \in [1,\dots,m]$:

\item If $\frac{|X_i|}{n}\le
c\frac{\rad_{x_i}(X_i)}{\Lambda}$ then $T_i =
\texttt{hierarchical-star-partition}(X_i,x_i,n,\Lambda)$;

\item
Otherwise, set $X_i$ to be a \textbf{reset cluster},
$T_i =
\texttt{hierarchical-star-partition}(X_i,x_i,|X_i|,\rad_{x_i}(X_i))$;

\item Let $T$ be the tree formed by connecting $T_0$ with $T_i$
using edge $(y_i,x_i)$ for each $i \in [1,\dots,m]$;
\end{enumerate}
\end{minipage}}\caption{\texttt{hierarchical-star-partition} algorithm} \label{fig:hirachical-tar}
\end{figure}
}

%
%***********************************************************
\subsection{Algorithm Analysis}
The hierarchical star-partition of $G=(X,E,\omega)$ naturally
induces a laminar family $\mathcal{F}\subseteq 2^X$. Let
$\mathcal{G}$ be the rooted \emph{construction tree} whose nodes
are sets in $\mathcal{F}$, $F\in\mathcal{F}$ is a parent of
$F'\in\mathcal{F}$ if $F'$ is a cluster formed by the partition of
$F$. Observe that the spanning tree $T$ obtained by our
hierarchical star decomposition has the property that every
$F\in\mathcal{F}$ corresponds to a sub tree $T[F]$ of $T$. Let
$\mathcal{R}\subseteq \mathcal{F}$ be the set of all reset
clusters. For each $F\in \mathcal{F}$, let $\mathcal{G}_F$ be the
sub-tree of the construction tree $\mathcal{G}$ rooted at $F$,
that contains all the nodes $X$ whose path to $F$ (excluding $F$
and $X$) contains no node in $\mathcal{R}$. For $F\in\mathcal{F}$
let $\mathcal{R}(F)\subseteq\mathcal{R}$ be the set of reset
cluster which are descendants of $F$ in $\mathcal{G}_F$ (These are
the leaves of the construction sub-tree $\mathcal{G}_F$ rooted at
$F$). In what follows we use the following convention on our
notation: whenever $X$ is a cluster in $\mathcal{G}$ with center
point $x_0$ with respect to which the star-partition of $X$ has
been constructed, we define $\rad(X) = \rad_{x_0}(X)$. We first
claim the following bound on $\alpha$ produced by the decompose
algorithms.
\begin{claim}\label{claim:diam-reduce}
Fix $F\in\mathcal{F}$ and $\mathcal{G}_F$. Let
$X\in\mathcal{G}_F\setminus \mathcal{R}(F)$, such that
$d_{\mathcal{G}}(X,F)=k$. By our construction, in each iteration of
the partition algorithm the radius decreases by a factor of at least
$\frac{5}{8}$, hence $\rad(X)\le\rad(F)\cdot(\frac{5}{8})^k$.
\end{claim}
\begin{proof}
For any cluster $F$, the radius of the central ball in the star
decomposition of $F$ is at most $((1/2) + \alpha) \rad(F)$. Since
the radius of this ball is also at least $(1/2) \rad(F)$ then the
radius of each cone is at most $((1/2) + \alpha) \rad(F)$ as well.
Let $Y\in\mathcal{R}$ such that $X\in\mathcal{G}_Y$. Since
$C=8\sqrt{c\cdot \hat{c}}$ then $\alpha = \sqrt{\epsilon_{\rm
lim}}/C =
\sqrt{\frac{|X|}{c|Y|}\left(\frac{\rad(Y)}{\rad(X)}\right)^{1/4}}/8
\le \frac{1}{8}\sqrt{\left(\frac{\rad(X)}{\rad(Y)}\right)^{3/4}} \le
\frac{1}{8}$.
\end{proof}
We now show that the spanning tree of each cluster increases its
diameter by at most a constant factor. Recall that $c' = e(2e+1)$.
\begin{lemma}\label{lem:small-radius}
For every $F\in\mathcal{F}$ and $T[F]\subseteq T$ we have
$\rad(T[F])\le c'\cdot \rad(F))$.
\end{lemma}
\begin{proof}
Let $Y\in\mathcal{R}$. We first prove by induction on the
construction tree $\mathcal{G}$ that for every $X\in\mathcal{G}_Y$
with $t=d_{\mathcal{G}}(X,Y)$ we have
\swapSODA{
\begin{eqnarray}
\rad(T[X])&\le&\prod_{j\ge
t}(1+\frac{1}{8}(\frac{7}{8})^j) \nonumber\\
&&\left(\rad(X)+\sum_{R\in \mathcal{R}(Y)\cap\mathcal{G}_X}\rad(T[R])\right)
\label{eq:rad-1}
\end{eqnarray}
}
{
\begin{equation}\label{eq:rad-1}
\rad(T[X])\le\prod_{j\ge
t}(1+\frac{1}{8}(\frac{7}{8})^j)\left(\rad(X)+\sum_{R\in
\mathcal{R}(Y)\cap\mathcal{G}_X}\rad(T[R])\right)
\end{equation}
}
\noSODA{
Fix some cluster $X\in\mathcal{G}_Y$, such that
$t=d_{\mathcal{G}}(X,Y)$ and assume the hypothesis is true for all
its children in $\mathcal{G}_Y$. If $X$ is a leaf of $\mathcal{G}_Y$
then it is a reset cluster and the claim trivially holds (since
$X\in\mathcal{R}(Y)\cap\mathcal{G}_X$). Otherwise, assume we
partition $X$ into $X_0,\dots,X_m$. Let $i\in[1,m]$ such that $X_i$
is the cluster such that $\omega(y_i,x_i) + \rad(T[X_i])$ is
maximal, hence $\rad(T[X]) \leq
\rad(T[X_0])+\omega(y_i,x_i)+\rad(T[X_i])$. There are four cases to
consider depending on whether $X_0$ and $X_i$ belong to
$\mathcal{R}$. Here we show the case of $X_0, X_i \not\in
\mathcal{R}$, the other cases are similar and easier.
}
Using \claimref{claim:diam-reduce} we obtain the following bound on
the increase in radius: $ \alpha\le
1/8\sqrt{\left(\frac{\rad(X)}{\rad(Y)}\right)^{3/4}}\le
1/8(5/8)^{3t/8}\le 1/8(7/8)^t$. It follows that
$\rad(X_0)+\omega(y_i,x_i)+\rad(X_i)\le\rad(X)(1+\alpha) \le
\rad(X)(1+1/8(7/8)^t)$.
\noSODA{
By the induction hypothesis we know that
$\rad(T[X_0])\le\prod_{j\ge
t+1}(1+\frac{1}{8}(\frac{7}{8})^j)(\rad(X_0)+\sum_{R\in
\mathcal{R}(Y)\cap\mathcal{G}_{X_0}}\rad(T[R]))$ and
$\rad(T[X_i])\le\prod_{j\ge
t+1}(1+\frac{1}{8}(\frac{7}{8})^j)(\rad(X_i)+\sum_{R\in
\mathcal{R}(Y)\cap\mathcal{G}_{X_i}}\rad(T[R]))$, hence
} \swapSODA { Then ~\eqref{eq:rad-1} can be proven by induction.} {
\begin{eqnarray*}
\rad(T[X]) &\leq
&\rad(T[X_0])+\omega(y_i,x_i)+\rad(T[X_i])\\
&\le &\prod_{j\ge
t+1}(1+\frac{1}{8}(\frac{7}{8})^j)\left(\rad(X_0)+\omega(y_i,x_i)+\rad(X_i)+
\sum_{R\in
\mathcal{R}(Y)\cap\mathcal{G}_X}\rad(T[R])\right)\\
&\le&\prod_{j\ge
t+1}(1+\frac{1}{8}(\frac{7}{8})^j)\left(\rad(X)(1+\frac{1}{8}(\frac{7}{8})^t)+\sum_{R\in
\mathcal{R}(Y)\cap\mathcal{G}_X}\rad(T[R])\right)\\
&\le&\prod_{j\ge
t}(1+\frac{1}{8}(\frac{7}{8})^j)\left(\rad(X)+\sum_{R\in
\mathcal{R}(Y)\cap\mathcal{G}_X}\rad(T[R])\right).
\end{eqnarray*}
This completes the proof of~\eqref{eq:rad-1}.
}
Now we continue to
prove the Lemma. First, we prove by induction on the construction
tree $\mathcal{G}$ that the Lemma holds for the set of reset
clusters. In fact we show a somewhat stronger bound. Recall that
$c=2e$. We show that for every cluster $Y\in\mathcal{R}$ we have
$\rad(T[Y])\le c\cdot\rad(Y)$. Assume the induction hypothesis is
true for all descendants of $Y$ in $\mathcal{R}$. In particular, for
all $R\in\mathcal{R}(Y)$, $\rad(T[R])\le c\cdot\rad(R)$. Recall that
$R$ becomes a reset cluster since
$\rad(R)\le\frac{\rad(Y)}{c\cdot|Y|}|R|$, hence
$\sum_{R\in\mathcal{R}(Y)}\rad(R)\le\rad(Y)/c$.
Using~\eqref{eq:rad-1} we have that \swapSODA{
\begin{eqnarray*}
\lefteqn{\rad(T[Y])}\\
&\le&\prod_{j\ge
0}(1+\frac{1}{8}(\frac{7}{8})^j)\left(\rad(Y)+\!\!\!\!\!\sum_{R\in
\mathcal{R}(Y)}\!\!\!\rad(T[R])\right)\\
&\le&(e^{\frac{1}{8}\sum_{j\ge 0}(\frac{7}{8})^j})(\rad(Y)+c\cdot\rad(Y)/c)\\
&\le&e\cdot 2\rad(Y)= c\cdot\rad(Y).
\end{eqnarray*}
}
{
\begin{eqnarray*}
\rad(T[Y])&\le&\prod_{j\ge
0}(1+\frac{1}{8}(\frac{7}{8})^j)\left(\rad(Y)+\sum_{R\in
\mathcal{R}(Y)}\rad(T[R])\right)\\
&\le&(e^{\frac{1}{8}\sum_{j\ge 0}(\frac{7}{8})^j})(\rad(Y)+c\cdot\rad(Y)/c)\\
&\le&e\cdot 2\rad(Y)= c\cdot\rad(Y).
\end{eqnarray*}
}
Finally, we show the Lemma holds for all the other clusters. Let
$F\in\mathcal{F}\setminus\mathcal{R}$ and $Y\in\mathcal{R}$ such
that $F\in\mathcal{G}_Y$. Let $t=d_{\mathcal{G}}(F,Y)$. Note that
$\sum_{R\in\mathcal{R}(Y)\cap\mathcal{G}_F}|R|=|F|$. Since
$F\notin\mathcal{R}$ we have
$\frac{\rad(Y)}{c|Y|}\le\frac{\rad(F)}{|F|}$ hence
\[
\sum_{R\in\mathcal{R}(Y)\cap\mathcal{G}_F}\rad(R)\le\frac{\rad(Y)}{c|Y|}
\sum_{R\in\mathcal{R}(Y)\cap\mathcal{G}_F}|R|\le\rad(F).
\]
By~\eqref{eq:rad-1} and the second induction \swapSODA{ it can be
shown that
\begin{eqnarray*}
\rad(T[F]) \le  e\cdot\rad(F)(c+1) = c'\cdot \rad(F),
\end{eqnarray*}
} {
we get \begin{eqnarray*} \rad(T[F])&\le&\prod_{j\ge
t}(1+\frac{1}{8}(\frac{7}{8})^j)
\left(\rad(F)+\sum_{R\in\mathcal{R}(Y)\cap\mathcal{G}_F}\rad(T[R])\right)\\
&\le&e\cdot\left(\rad(F)+c\sum_{R\in\mathcal{R}(Y)\cap\mathcal{G}_F}\rad(R)\right)\\
&\le&e\cdot\rad(F)(c+1) = c'\cdot \rad(F),
\end{eqnarray*}
}
proving the Lemma.
\end{proof}
We now proceed to bound for every $\epsilon$ the number of pairs
with distortion $\Omega(\sqrt{1/\epsilon})$, thus proving the
scaling distortion of our constructed the spanning tree. We begin
with some definitions that will be crucial in the analysis.
\begin{definition}
For each $\epsilon\in(0,1]$ and $R\in\mathcal{R}$ let
$\mathcal{K}(R,\epsilon)=\{F\in\mathcal{G}_R\mid
|F|<\epsilon/\hat{c} \cdot |R|\}$.
\end{definition}
Hence, a cluster is in $\mathcal{K}(R,\epsilon)$ if it contains
less than $\epsilon/\hat{c}$ fraction of the points of $R$.
Informally, when counting the badly distorted edges for a given
$\epsilon$, whenever we reach a cluster in
$\mathcal{K}(R,\epsilon)$ we count all its pairs as bad. If
$X\in\mathcal{G}_R$ then let
$\mathcal{K}(X,\epsilon)=\mathcal{K}(R,\epsilon)\cap\mathcal{G}_X$.
For $R\in\mathcal{R}$ let $\mathcal{G}_{R,\epsilon}$ be the
sub-tree rooted at $R$, that contains all the nodes $X$ whose path
to $R$ (excluding $R$ and $X$) contains no node in
$\mathcal{R}\cup\mathcal{K}(R,\epsilon)$. Observe that
$\mathcal{G}_{R,\epsilon}$ is a sub tree of $\mathcal{G}_R$.
\begin{comment}  %% removed due to the new parameter \hat{c}.
In the following lemma we bound the number of pairs with
distortion $\Omega(\sqrt{1/\epsilon})$. For every $\epsilon$ we
will bound $B_\epsilon(X)$ by $C'\cdot\epsilon {n\choose 2}$. By
letting $\epsilon' = C' \cdot\epsilon$ and allowing the distortion
to increase by a $\sqrt{C'}$ factor we get the scaling distortion
bound.
\end{comment}
\begin{lemma}\label{lem:few-bad}
For any $R\in\mathcal{R}$, $\epsilon\in(0,1]$ we have that
$B_\epsilon(R)\le \epsilon {R\choose 2}$.
\end{lemma}
\begin{proof}
Fix some $\epsilon\in(0,1]$. Fix $F\in\mathcal{R}$. In order to
prove the claim for $F$, we will first prove the following
inductive claim for all $X\in\mathcal{G}_F$. Let
$t=d_{\mathcal{G}}(X,F)$. Let $\mathcal{E}(X)=\left({X\choose
2}\setminus{\bigcup_{R\in\mathcal{R}(X)}{R\choose 2}\cup
\bigcup_{K\in\mathcal{K}(X,\epsilon)}{K\choose 2}}\right)$.
\swapSODA{
\begin{eqnarray}
B_\epsilon(X)&\le&\frac{2}{\hat{c}}\cdot
\epsilon\sum_{i\ge t}(9/10)^i\cdot|\mathcal{E}(X)| \nonumber\\
&& \,+
\sum_{R\in\mathcal{R}(F)\cap\mathcal{G}_X} B_\epsilon(R) +
\sum_{K\in\mathcal{K}(F,\epsilon)\cap\mathcal{G}_X} B_\epsilon(K).
\label{eq:bad-eps-1}
\end{eqnarray}
}
{
\begin{equation}
\label{eq:bad-eps-1} B_\epsilon(X)\le \frac{2}{\hat{c}}\cdot
\epsilon\sum_{i\ge t}(9/10)^i\cdot|\mathcal{E}(X)| +
\sum_{R\in\mathcal{R}(F)\cap\mathcal{G}_X} B_\epsilon(R) +
\sum_{K\in\mathcal{K}(F,\epsilon)\cap\mathcal{G}_X} B_\epsilon(K).
\end{equation}
}
The base of the induction, where $X$ is a leaf in $\mathcal{G}_F$,
 \ie  $X\in\mathcal{R}(F)\cup\mathcal{K}(F,\epsilon)$, is
trivial. Assume the claim holds for all the children $X_0,\dots,X_m$
of $X$. Let $P= \{P_i\}_{i=0}^m$ be the star-partition of $X$, where
$P_i = (X_i,Y_i)$, $Y_i=\cup_{j=i+1}^m X_j$. Recall the definition
of $\hat{B}_\epsilon(P_i)=|\{(x,y)\mid x\in X_i\wedge y\in Y_i\wedge
d(x,y) \le\frac{\sqrt{\epsilon}\cdot\hat{\Lambda}}{150C}\}|$, where
$\hat{\Lambda}=\rad(X)$. Denote
$\hat{B}_\epsilon(P)=\sum_{i=0}^m\hat{B}_\epsilon(P_i)$.  By
\lemmaref{lem:small-radius} we have that $\rad(T(X))\le c'\rad(X)$.
Hence, the number of pairs distorted more than $150C\cdot
c'\sqrt{1/\epsilon}$ by the partition $P$ is bounded by
$\hat{B}_\epsilon(P)$. Now, since $X \notin \mathcal{K}(F,\epsilon)$
then $\epsilon < \hat{c}\cdot |X|/|F| \leq 1/\beta\cdot |X|/|F| =
\epsilon_{\rm lim}$. Hence we can apply \lemmaref{lem:partition} to
deduce a bound on $B_\epsilon(P_i)$. By \claimref{claim:diam-reduce}
we have $\beta=
\frac{1}{\hat{c}}\left(\frac{\rad(X)}{\rad(F)}\right)^{1/4}
\le\frac{1}{\hat{c}} (\frac{5}{8})^{t/4}$. From
\lemmaref{lem:partition} we obtain \swapSODA{
\begin{eqnarray*}
\hat{B}_\epsilon(P)&=&\sum_{i=0}^m\hat{B}_\epsilon(P_i)\le
\frac{1}{\hat{c}}\cdot
\epsilon(\frac{5}{8})^{t/4}\sum_{i=0}^m|X_i||F\setminus X_i|\\
&\le&\frac{2}{\hat{c}}\cdot \epsilon(9/10)^t|\mathcal{E}(X)|.
\end{eqnarray*}
}
{
\[
\hat{B}_\epsilon(P)=\sum_{i=0}^m\hat{B}_\epsilon(P_i)\le
\frac{1}{\hat{c}}\cdot
\epsilon(\frac{5}{8})^{t/4}\sum_{i=0}^m|X_i||F\setminus X_i| \le
\frac{2}{\hat{c}}\cdot \epsilon(9/10)^t|\mathcal{E}(X)|.
\]
} Using the induction hypothesis \swapSODA{it can be shown that the
inductive claim holds.
} {
we get that \begin{eqnarray*}
B_\epsilon(X)&\le&\hat{B}_\epsilon(P) +\sum_{j=0}^mB_\epsilon(X_j)\\
&\le& \frac{2}{\hat{c}}\cdot \epsilon(9/10)^t|\mathcal{E}(X)| +
\sum_{j=0}^m\left(\frac{2}{\hat{c}}\cdot
\epsilon|\mathcal{E}(X_j)|\sum_{i\ge t+1}(9/10)^i +
\sum_{R\in\mathcal{R}(F)\cap\mathcal{G}_{X_j}}B_\epsilon(R) +
\sum_{K\in\mathcal{K}(F,\epsilon)\cap\mathcal{G}_{X_j}}B_\epsilon(K)\right)\\
&\le &\frac{2}{\hat{c}}\cdot \epsilon(9/10)^t|\mathcal{E}(X)| +
\frac{2}{\hat{c}}\cdot \epsilon|\mathcal{E}(X)|\sum_{i\ge
t+1}(9/10)^i +
\sum_{R\in\mathcal{R}(F)\cap\mathcal{G}_X}B_\epsilon(R) +
\sum_{K\in\mathcal{K}(F,\epsilon)\cap\mathcal{G}_X}B_\epsilon(K)\\
&\le& \frac{2}{\hat{c}}\cdot \epsilon\sum_{i\ge
t}(9/10)^i|\mathcal{E}(X)| +
\sum_{R\in\mathcal{R}(F)\cap\mathcal{G}_X}B_\epsilon(R) +
\sum_{K\in\mathcal{K}(F,\epsilon)\cap\mathcal{G}_X}B_\epsilon(K),
\end{eqnarray*}
which proves the inductive claim.
}
We now prove the Lemma by induction on the construction tree
$\mathcal{G}$. Let $F\in\mathcal{R}$. By the induction hypothesis
$B_\epsilon(R)\le \epsilon {R \choose 2}$ for every
$R\in\mathcal{R}(F)$. Observe that if $K\in \mathcal{K}(F,\epsilon)$
then we discard all pairs in $K$. Hence $B_\epsilon(K)\le |K|^2\le
\frac{1}{\hat{c}}\cdot \epsilon|F|\cdot|K|$. Recall that $\hat{c} =
22$. From~\eqref{eq:bad-eps-1} we obtain \swapSODA{
\begin{eqnarray*}
\lefteqn{B_\epsilon(F)}\\
&\le& \frac{2}{22}\cdot \epsilon\sum_{i\ge
0}(9/10)^i\cdot|\mathcal{E}(F)| +
\epsilon\sum_{R\in\mathcal{R}(F)}{R\choose 2}\\
&& +\sum_{K\in\mathcal{K}(F,\epsilon)} \frac{1}{22}\cdot \epsilon|F|\cdot|K|\\
&\le& \left[\frac{20}{22}\cdot \epsilon\cdot|\mathcal{E}(F)| +
\frac{20}{22}\epsilon\sum_{R\in\mathcal{R}(F)}{R\choose 2}\right]\\
&& + \left[\frac{2}{22}\epsilon\sum_{R\in\mathcal{R}(F)}\!\!\!|R|\frac{|R|-1}{2}+\frac{1}{22}
\epsilon|F|\sum_{K\in\mathcal{K}(F,\epsilon)}|K|\right]\\
&\le& \frac{20}{22}\epsilon{|F|\choose 2} + \frac{1}{22}
\epsilon|F|\left(\sum_{R\in\mathcal{R}(F)}\!\!\!|R|\!-\!1\!
+\!\!\!\!\!\sum_{K\in\mathcal{K}(F,\epsilon)}\!\!\!\!|K|\right)\\
&\le& \frac{20}{22}\epsilon{|F|\choose 2} + \frac{1}{22}
\epsilon\cdot
|F|(|F|-1)=\epsilon{|F|\choose 2},
\end{eqnarray*}
}
{
\begin{eqnarray*}
B_\epsilon(F)&\le& \frac{2}{22}\cdot \epsilon\sum_{i\ge
0}(9/10)^i\cdot|\mathcal{E}(F)| +
\epsilon\sum_{R\in\mathcal{R}(F)}{R\choose 2} +
\sum_{K\in\mathcal{K}(F,\epsilon)} \frac{1}{22}\cdot \epsilon|F|\cdot|K|\\
&\le& \left[\frac{20}{22}\cdot \epsilon\cdot|\mathcal{E}(F)| +
\frac{20}{22}\epsilon\sum_{R\in\mathcal{R}(F)}{R\choose 2}\right]
+ \left[\frac{2}{22}
\epsilon\sum_{R\in\mathcal{R}(F)}|R|\cdot(|R|-1)/2 + \frac{1}{22}
\epsilon\cdot|F|\sum_{K\in\mathcal{K}(F,\epsilon)}|K|\right]\\
&\le& \frac{20}{22} \epsilon{|F|\choose 2} + \frac{1}{22}
\epsilon\cdot |F|\left(\sum_{R\in\mathcal{R}(F)}(|R|-1) +
\sum_{K\in\mathcal{K}(F,\epsilon)}|K|\right)\\
&\le& \frac{20}{22}\epsilon{|F|\choose 2} + \frac{1}{22}
\epsilon\cdot
|F|(|F|-1)\\
&=&\epsilon{|F|\choose 2},
\end{eqnarray*}
}
where the third inequality follows from the definition of
$\mathcal{E}(X)$ and from the fact that for each
$K\in\mathcal{K}(F,\epsilon)$, $R\in\mathcal{R}(F)$ we have $K\cap
R=\emptyset$.
\end{proof}
Applying \lemmaref{lem:few-bad} on the original graph proves
\theoremref{thm:scaling-spanning-trees}. Finally, we complete the
proof of \lemmaref{lem:partition} stating the properties of our
generic decompose algorithm.
\begin{proof}[Proof of \lemmaref{lem:partition}]
\begin{comment}
First we handle the situation where
$B(u,(\theta+\alpha)\cdot\hat{\Lambda})\setminus
B(u,(\theta+(\frac{7}{8})\alpha)\cdot\hat{\Lambda})=\emptyset$.
Setting $r=(\theta+\alpha)\cdot\hat{\Lambda}$ and defining
$Z=B(u,r)$ will satisfy for all $\epsilon \in (0,\epsilon_{\rm
lim}]$ : $\hat{B}_\epsilon(P)=0$, since for all $x\in Z$, $y\in
\bar{Z}$ we have
$d(x,y)\ge (\frac{1}{8})\alpha\cdot\hat{\Lambda}\ge\sqrt{\epsilon}\hat{\Lambda}$.\\
Let $x\in B(u,(\theta+\alpha)\cdot\hat{\Lambda})\setminus
B(u,(\theta+(\frac{7}{8})\alpha)\cdot\hat{\Lambda})$, let
$t=d(u,x)$.
\end{comment}
We distinguish between the following two cases:
\begin{description}
\item [Case 1:] $|B(u,(\theta +\alpha/2)\hat{\Lambda})|\le n/2$.
In this case let $\hat{\epsilon}=\max\{\epsilon\in(0,
\epsilon_{\rm lim}] \mid
|B(u,(\theta+\frac{\sqrt{\epsilon}}{4C})\hat{\Lambda})|
\ge\epsilon\cdot\beta\cdot n\}$. Let
$\hat{S}=[(\theta+\frac{\sqrt{\epsilon}}{4C})\hat{\Lambda},(\theta+\frac{\sqrt{\epsilon}}{2C})\hat{\Lambda})$,
and
$S=\left[\left(\theta+\frac{\sqrt{\epsilon}}{C}\left(\frac{1}{4}+\frac{1}{25}\right)\right)\hat{\Lambda},
\left(\theta+\frac{\sqrt{\epsilon}}{C}\left(\frac{1}{2}-\frac{1}{25}\right)\right)\hat{\Lambda}\right]$.
\item [Case 2:] $|B(u,(\theta +\alpha/2)\hat{\Lambda})| > n/2$. In
this case let $\hat{\epsilon}=\max\{ \epsilon\in[0, \epsilon_{\rm
lim}] \mid |W \setminus
B(u,(\theta+\alpha-\frac{\sqrt{\epsilon}}{4C})\hat{\Lambda})|
\ge\epsilon\cdot\beta\cdot n\}$. Let
$\hat{S}=[(\theta+\alpha-\frac{\sqrt{\epsilon}}{2C})\hat{\Lambda},
(\theta+\alpha-\frac{\sqrt{\epsilon}}{4C})\hat{\Lambda}]$, and
$S=\left[\left(\theta\!+\!\alpha\!-\!\frac{\sqrt{\epsilon}}{C}
\left(\frac{1}{2}-\frac{1}{25}\right)\right)\hat{\Lambda},
\left(\theta\!+\!\alpha\!-\!\frac{\sqrt{\epsilon}}{C}
\left(\frac{1}{4}+\frac{1}{25}\right)\right)\hat{\Lambda}\right]$.
%(Note that in this case it may be that $\hat{\epsilon}=0$, which
%does not interfere with the proof.)
\end{description}
%Notice that the interval $\hat{S}$ of Case 1 is disjoint from the
%interval $\hat{S}$ of Case 2.
We show that one can choose $r\in S$ and define $Z=B(u,r)$ such
that the property of the Lemma holds.
We now show the property of the Lemma holds for all
$\epsilon\in(32\hat{\epsilon},\epsilon_{\rm lim}]$ and any $r\in
S$.
 \begin{description}
\item[Proof for Case 1:] In this case we will use the bound:
\begin{equation}
\label{eq:bad-bound-big-eps-1} \hat{B}_\epsilon(P) \le
|B(u,r+\sqrt{\epsilon}\hat{\Lambda}/(150C))\setminus Z|\cdot|Z|.
\end{equation}
Note that
$r+\sqrt{\epsilon}\hat{\Lambda}/(150C)\le(\theta+\sqrt{\hat{\epsilon}}/(2C))\hat{\Lambda}+
\sqrt{\epsilon/2}\hat{\Lambda}/(8C)
\le(\theta+\sqrt{\epsilon/2}/(4C))\hat{\Lambda}$, using that
$\hat{\epsilon}\le \epsilon/32$. Now, by the maximality of
$\hat{\epsilon}$ we have
$|B(u,(\theta+\sqrt{\epsilon/2}/(4C))\hat{\Lambda})|\le\epsilon/2\cdot\beta\cdot
n$. Therefore, using~\eqref{eq:bad-bound-big-eps-1} we get
\begin{eqnarray*}
\hat{B}_\epsilon(P) &\le&
|B(u,(\theta+\sqrt{\epsilon/2}/(4C)\hat{\Lambda})|\cdot|Z|\\
&\le& (\epsilon\cdot\beta\cdot n/2)\cdot|Z| \le
\epsilon\cdot\beta\cdot|Z|\cdot(n-|Z|),
\end{eqnarray*}
using $|Z|\le n/2$.

\swapSODA{\end{description} The proof of the second case appears
in the full version.} { \item [Proof for Case 2:] In this case we
will use the bound:
\begin{eqnarray}
\label{eq:bad-bound-big-eps-2} \hat{B}_\epsilon(P) &\le& |\bar{Z}|
\cdot |W\setminus B(u,r-\sqrt{\epsilon}\hat{\Lambda}/(150C))|.
\end{eqnarray}
Note that $r - \sqrt{\epsilon}\hat{\Lambda}/(150C) \ge
(\theta+\alpha-\sqrt{\hat{\epsilon}}/(2C))\hat{\Lambda} -
\sqrt{\epsilon/2}\hat{\Lambda}/(8C) \geq
(\theta+\alpha-\sqrt{\epsilon/2}\hat{\Lambda}/(4C)
)\hat{\Lambda}$, using that $\hat{\epsilon}\le \epsilon/32$. Now,
from the maximality of $\hat{\epsilon}$ we have $|W \setminus
B(u,(\theta+\alpha-\sqrt{\epsilon/2}/(4C))\hat{\Lambda})|<\epsilon\cdot\beta\cdot
n/2$. Therefore, using~\eqref{eq:bad-bound-big-eps-2} we get
\begin{eqnarray*}
\hat{B}_\epsilon(P)&\le& |\bar{Z}|\cdot |W \setminus B(u,(\theta+\alpha-\sqrt{\epsilon/2}/(4C))\hat{\Lambda})|\\
&\le& |\bar{Z}| \cdot \epsilon\cdot\beta\cdot n/2 \leq
\epsilon\cdot\beta\cdot |Z|(n-|Z|),
\end{eqnarray*}
using $|Z|\ge n/2$.
\end{description}
}

We next show the property of the Lemma hold for all
$\epsilon\in(0,32\hat{\epsilon}]$. We will prove the claim for
Case 1. The argument for Case 2 is the analogous.
As before we define $Q=\{w\mid d(u,w)\in \hat{S}\}$. Now we have
\begin{claim} \label{claim:small-x1-variation}
$|Q| \leq 4\cdot\hat{\epsilon}\cdot\beta\cdot n$.
\end{claim}
\begin{proof}
We have $Q \subseteq B(u,(\theta+
\sqrt{\hat{\epsilon}}/(2C))\hat{\Lambda})$. We distinguish between
2 cases: If $\hat{\epsilon} \leq \epsilon_{\rm lim}/4$ then
$|B(u,(\theta+\sqrt{4\hat{\epsilon}}/(4C))\hat{\Lambda})| \leq
4\hat{\epsilon}\cdot\beta\cdot n$ (by the maximality of
$\hat{\epsilon}$). Otherwise, $\hat{\epsilon}\in (\epsilon_{\rm
lim}/4,\epsilon_{\rm lim}]$. In this case $|Q| \le |W| \leq
\epsilon_{\rm lim}\cdot\beta\cdot n\le
4\hat{\epsilon}\cdot\beta\cdot n$.
\end{proof}
As before we will choose some $r\in S$ and the partition $P$ will
be $Z=B(u,r)$, $\bar{Z}=W\setminus Z$. It is easy to check that
for any $r\in S$ we get $\hat{\epsilon}\cdot n\cdot\beta\le |Z|\le
n/2$. We now find $r\in S$ which satisfy the property of the Lemma
for all $0<\epsilon\le 32\hat{\epsilon}$: For any $r \in S$ and
$\epsilon \leq 32\hat{\epsilon}$ let
$S_r(\epsilon)=[r-\sqrt{\epsilon}\hat{\Lambda}/(150C)),r+\sqrt{\epsilon}\hat{\Lambda}/(150C))]$,
$s(\epsilon)=\sqrt{\epsilon}\hat{\Lambda}/(75C)$ and let
$Q_r(\epsilon)=\{w \mid d(u,w)\in S_r(\epsilon)\}$. Note that the
length of the interval $S$ is given by
$s=17/(100C)\sqrt{\hat{\epsilon}}\hat{\Lambda}$. We say that
properly $A_r(\epsilon)$ holds if cutting at radius $r$ is ``good"
for $\epsilon$, formally: $A_r(\epsilon)$ iff $|Q_r(\epsilon)|
\le\sqrt{\epsilon\cdot\hat{\epsilon}/2}\cdot n\cdot\beta$. Notice
that only pairs $(x,y)$ such that $x,y\in Q_r(\epsilon)$ may be
distorted by more than $150C\sqrt{1/\epsilon}$.
\begin{claim}
\label{claim:small-eps-property-beta}
 There exists some $r\in S$ such that properly $A_r(\epsilon)$
holds for all $\epsilon \in (0, 32\hat{\epsilon}]$.
\end{claim}

\noSODA{
\begin{proof}
As the proof of \claimref{claim:small-eps} goes, we conduct exactly
the same iterative process that greedily deletes the ``worst"
interval in $S$, which are $\{S_{r_j}(\epsilon_j)\}_{j=1}^t$, and we
remain with $I_t\subseteq S$. We now argue that $I_t\neq\emptyset$.
As before we have $ \sum_{j=1}^t|Q_{r_j}(\epsilon_j)| \le 2|Q|\le
8\hat{\epsilon}\cdot\beta\cdot n$. Recall that since
$A_{r_j}(\epsilon_j)$ does not hold then for any $1\le j\le t$ :
$|Q_{r_j}(\epsilon_j)|
> \sqrt{\epsilon_j\cdot\hat{\epsilon}/2}\cdot\beta\cdot n$ which implies that
$
\sum_{j=1}^t\sqrt{\epsilon_j} < 12\sqrt{\hat{\epsilon}}.
$
On the other hand, by definition
\[
\sum_{j=1}^t s(\epsilon_j)\le
\sum_{j=1}^t\sqrt{\epsilon_j}\Delta/(75C)\le
12/(75C)\cdot\sqrt{\hat{\epsilon}}\Delta =
16/(100C)\cdot\sqrt{\hat{\epsilon}}\Delta.
\]
Since $s=17/(100C)\cdot \sqrt{\hat{\epsilon}}\Delta$ then indeed
$I_t\neq\emptyset$ so any $r\in I_t$ satisfies the condition of
the claim.
\end{proof}
} % end of noSODA

\inSODA{The proof appears in the full version.}
\claimref{claim:small-eps-property-beta} shows that for any
$\epsilon\in(0,32\hat{\epsilon}]$ we have
\[
\hat{B}_\epsilon(P)\le\epsilon\cdot\hat{\epsilon}/2 \cdot
(n\cdot\beta)^2\le \epsilon\cdot\beta\cdot|Z|\cdot(n-|Z|),
\]
which concludes the proof of the lemma.
\end{proof}

\section{Probabilistic Scaling Embedding into spanning trees}
The proof of this theorem is based on a somewhat simpler variation
of the decomposition algorithm from the previous section. In fact,
the \texttt{hierarchical-star-partition} algorithm remains
practically the same, with modified sub-method
\texttt{probabilistic-star-partition} given in
\figureref{fig:scaling-star}, instead of \texttt{star-partition}.

Let $f:\mathbb{R}\rightarrow\mathbb{R}_+$ be a monotone
non-decreasing function satisfying
\begin{equation} \label{eqn:f}
\int_1^{\infty}\frac{dx}{f(x)}=1~.
\end{equation}
For example if we define
$\log^{(0)}n=n$, and for any $i>0$ define recursively
$\log^{(i)}n=\log(\log^{(i-1)}n)$, then we can take for any
constants $\theta>0$, $t\in\mathbb{N}$ the function $
f(n)=\hat{c}\prod_{j=0}^{t-1}\log^{(j)}(n)\cdot\left(\log^{(t)}(n)\right)^{1+\theta}$,
for sufficiently small constant $\hat{c}>0$, and it will satisfy the
conditions.

\begin{figure}[ht!]
\fbox{
\begin{minipage}[t]{180mm}
$(X_0,\dots,X_t,(y_1,x_1),\dots,(y_t,x_t)) =
\texttt{probabilistic-star-partition}(X,x_0,\Lambda)$:
\begin{enumerate}
\item Set $k=0$ ; $\hat{\Lambda}=\rad_{x_0}(X)$; $\alpha =
\frac{1}{f(\log(2\Lambda/\hat{\Lambda}))}$;

\item Choose uniformly at random $\beta\in [0,1/8]$.

\item Let $\gamma$ be the value in $\{0,1/16\}$ minimizing
$|B(x_0,(1/2+\gamma+1/16)\hat{\Lambda})|-
|B(x_0,(1/2+\gamma)\hat{\Lambda})|$.

\item $X_0=B(x_0,(1/2+3\gamma/2+\beta/4)\hat{\Lambda})$; $Y_0=X\setminus X_0$;

\item If $Y_k = \emptyset$ set $t=k$ and stop; Otherwise, set
$k=k+1$;

\item Let $v_k\in Y_{k-1}$ be the point minimizing $\hat{\chi}_k=\frac{|Y_0|}{|B_{Y_0}(x,\alpha\hat{\Lambda}/64)|}$;
Set $\chi_k=\max\{4,\hat{\chi}_k\}$;

\item Choose $r\in [\alpha\hat{\Lambda}/16,\alpha\hat{\Lambda}/8]$
according to the distribution
$p(r)=\frac{\chi_k^2}{1-\chi_k^{-2}}\frac{32\ln\chi_k}{\alpha\hat{\Lambda}}\cdot\chi_k^{-32r/(\alpha\hat{\Lambda})}$;

\item Let $(x_k,y_k)$ be the edge in $E$ which lies on a shortest path from $v_k$ to $x_0$
such that $y_k\in X_0,x_k\in Y_{k-1}$\footnote{By the definition of
cone-metric, if $z_k\in Y_{k-1}$ all the points on any shortest path
from $v_i$ to $x_0$ are either in $X_0$ or in $Y_{k-1}$};

\item Let $\ell=\ell^{x_0}_{x_k}$ be the cone-metric with respect
to $x_0$ and $x_k$ on the subspace $Y_{k-1}$;\\
$X_k=B_{(Y_{k-1},\ell)}(x_k,r)$; $Y_k=Y_{k-1}\setminus X_k$.

\item goto 4;
\end{enumerate}
\end{minipage}}\caption{\texttt{probabilistic-star-partition} algorithm} \label{fig:scaling-star}
\end{figure}

\subsection{Algorithm Analysis}
Let $\hat{\mathcal{H}}$ be the distribution on laminar families
induced by the algorithm above. Let $\mathcal{H} =
supp(\hat{\mathcal{H}})$. We have the following analogs of
\claimref{claim:diam-reduce} and \lemmaref{lem:small-radius}.

\begin{claim}\label{claim:prob-diam-reduce}
Fix $\mathcal{F} \in \mathcal{H}$, $F\in\mathcal{F}$. Let
$X\in\mathcal{G}_F\setminus \mathcal{R}(F)$, such that
$d_{\mathcal{G}}(X,F)=k$. By our construction, in each iteration
of the partition algorithm the radius decreases by a factor of at
least $5/8$. Hence
$$
\rad(X)\le\rad(F)\cdot(5/8)^k~.
$$
\end{claim}

\noSODA{
\begin{proof}
For any cluster $F$, the radius of the central ball in the star
decomposition of $F$ is at most $(5/8)\rad(F)$. Since the radius of
this ball is also at least $(1/2) \rad(F)$ then the radius of each
cone is at most $((1/2) + \alpha/8) \rad(F)\le (5/8)\rad(F)$ as
well.
\end{proof}

We now show that the spanning tree of each cluster increases its
diameter by at most a constant factor. Recall that $c' = e(2e+1)$.

}%end of noSODA

\begin{lemma}\label{lem:prob-small-radius}
For every $\mathcal{F} \in \mathcal{H}$, $F\in\mathcal{F}$ we have
$\rad(T[F])\le c'\cdot \rad(F)$.
\end{lemma}

\noSODA{
\begin{proof}
Let $Y\in\mathcal{R}$. We first prove by induction on the
construction tree $\mathcal{G}$ that for every $X\in\mathcal{G}_Y$
with $t=d_{\mathcal{G}}(X,Y)$ we have
\begin{equation}\label{eq:prob-rad-1}
\rad(T[X])\le\prod_{j\ge t}(1+1/(8f(1+j/5)))\left(\rad(X)+\sum_{R\in
\mathcal{R}(Y)\cap\mathcal{G}_X}\rad(T[R])\right)
\end{equation}
Fix some cluster $X\in\mathcal{G}_Y$, such that
$t=d_{\mathcal{G}}(X,Y)$ and assume the hypothesis is true for all
its children in $\mathcal{G}_Y$. If $X$ is a leaf of $\mathcal{G}_Y$
then it is a reset cluster and the claim trivially holds (since
$X\in\mathcal{R}(Y)\cap\mathcal{G}_X$). Otherwise, assume we
partition $X$ into $X_0,\dots,X_m$. Let $i\in[1,m]$ such that $X_i$
is the cluster such that $\omega(y_i,x_i) + \rad(T[X_i])$ is
maximal, hence $\rad(T[X]) \leq
\rad(T[X_0])+\omega(y_i,x_i)+\rad(T[X_i])$. There are four cases to
consider depending on whether $X_0$ and $X_i$ belong to
$\mathcal{R}$. Here we show the case of $X_0, X_i \not\in
\mathcal{R}$, the other cases are similar and easier. Using
\claimref{claim:prob-diam-reduce} $\log(2\rad(Y)/\rad(X))\ge 1+t/5$,
it follows that
\[
\rad(X_0)+\omega(y_i,x_i)+\rad(X_i)\le\rad(X)\left(1+
1/(8f(\log(2\rad(Y)/\rad(X))))\right) \le
\rad(X)\left(1+1/(8f(1+t/5))\right)
\]
By the induction hypothesis we know that $\rad(T[X_0])\le\prod_{j\ge
t+1}(1+1/(8f(1+j/5)))(\rad(X_0)+\sum_{R\in
\mathcal{R}(Y)\cap\mathcal{G}_{X_0}}\rad(T[R]))$ and
$\rad(T[X_i])\le\prod_{j\ge
t+1}(1+1/(8f(1+j/5)))(\rad(X_i)+\sum_{R\in
\mathcal{R}(Y)\cap\mathcal{G}_{X_i}}\rad(T[R]))$, hence
\begin{eqnarray*}
\rad(T[X]) &\leq
&\rad(T[X_0])+\omega(y_i,x_i)+\rad(T[X_i])\\
&\le &\prod_{j\ge
t+1}(1+1/(8f(1+j/5)))\left(\rad(X_0)+\omega(y_i,x_i)+\rad(X_i)+
\sum_{R\in
\mathcal{R}(Y)\cap\mathcal{G}_X}\rad(T[R])\right)\\
&\le&\prod_{j\ge
t+1}(1+1/(8f(1+j/5)))\left(\rad(X)(1+1/(8f(1+t/5)))+\sum_{R\in
\mathcal{R}(Y)\cap\mathcal{G}_X}\rad(T[R])\right)\\
&\le&\prod_{j\ge t}(1+1/(8f(1+j/5)))\left(\rad(X)+\sum_{R\in
\mathcal{R}(Y)\cap\mathcal{G}_X}\rad(T[R])\right).
\end{eqnarray*}

This completes the proof of~\eqref{eq:prob-rad-1}. Now we continue
to prove the Lemma. First, we prove by induction on the construction
tree $\mathcal{G}$ that the Lemma holds for the set of reset
clusters. In fact we show a somewhat stronger bound. Recall that
$c=2e$. We show that for every cluster $Y\in\mathcal{R}$ we have
$\rad(T[Y])\le c\cdot\rad(Y)$. Assume the induction hypothesis is
true for all descendants of $Y$ in $\mathcal{R}$. In particular, for
all $R\in\mathcal{R}(Y)$, $\rad(T[R])\le c\cdot\rad(R)$. Recall that
$R$ becomes a reset cluster since
$\rad(R)\le\frac{\rad(Y)}{c\cdot|Y|}|R|$, hence
$\sum_{R\in\mathcal{R}(Y)}\rad(R)\le\rad(Y)/c$. Using
\eqnref{eq:prob-rad-1} and then \eqnref{eqn:f}, we have that
\begin{eqnarray*}
\rad(T[Y])&\le&\prod_{j\ge
0}(1+1/(8f(1+j/5)))\left(\rad(Y)+\sum_{R\in
\mathcal{R}(Y)}\rad(T[R])\right)\\
&\le&(e^{1/8\sum_{j\ge 0}1/f(1+j/5)})(\rad(Y)+c\cdot\rad(Y)/c)\\
&\le&e^{5/8}\cdot 2\rad(Y)=\le c\cdot\rad(Y).
\end{eqnarray*}

Finally, we show the Lemma holds for all the other clusters. Let
$F\in\mathcal{F}\setminus\mathcal{R}$ and $Y\in\mathcal{R}$ such
that $F\in\mathcal{G}_Y$. Let $t=d_{\mathcal{G}}(F,Y)$. Note that
$\sum_{R\in\mathcal{R}(Y)\cap\mathcal{G}_F}|R|=|F|$. Since
$F\notin\mathcal{R}$ we have
$\frac{\rad(Y)}{c|Y|}\le\frac{\rad(F)}{|F|}$ hence
\[
\sum_{R\in\mathcal{R}(Y)\cap\mathcal{G}_F}\rad(R)\le\frac{\rad(Y)}{c|Y|}
\sum_{R\in\mathcal{R}(Y)\cap\mathcal{G}_F}|R|\le\rad(F).
\]
By~\eqref{eq:prob-rad-1} and the second induction we get
\begin{eqnarray*}
\rad(T[F])&\le&\prod_{j\ge t}(1+1/(8f(j/5)))
\left(\rad(F)+\sum_{R\in\mathcal{R}(Y)\cap\mathcal{G}_F}\rad(T[R])\right)\\
&\le&e\cdot\left(\rad(F)+c\sum_{R\in\mathcal{R}(Y)\cap\mathcal{G}_F}\rad(R)\right)\\
&\le&e\cdot\rad(F)(c+1) = c'\cdot \rad(F),
\end{eqnarray*}
proving the Lemma.
\end{proof}

}%end of noSODA

For any $i>0$ let $\hat{\mathcal{H}}^{(i)}$ be the distribution on
laminar families induced by $i$ iterations of our probabilistic
\texttt{hierarchical-star-partition} algorithm. Let
$\mathcal{H}^{(i)} = supp(\hat{\mathcal{H}}^{(i)})$. Given
$\mathcal{F}^{(i)} \in \mathcal{H}^{(i)}$. Let $\mathcal{G}^{(i)}$
be the corresponding construction tree of $\mathcal{F}^{(i)}$. Given
$\mathcal{F}^{(i)}$, for any $x \in X$ let $F_i(x)$ be the leaf in
$\mathcal{G}^{(i)}$ containing $x$.

Given $x,y \in G$ and $j>0$ define events  $\mathcal{C},
\mathcal{C}_{\textrm{ball}}, \mathcal{X},\mathcal{Y},\mathcal{Z}$ as
follows:

\begin{itemize}

\item
Let $\mathcal{C}(x,y,j)$ be the event that there exists $i>0$
and $\mathcal{F}^{(i)} \in \mathcal{H}^{(i)}$ such that the
following holds:
\begin{enumerate}

\item $(\frac{8}{5})^j\le\rad(F_i(x))<(8/5)^{j+1}$.

\item $B(x,d(x,y))\subseteq F_{i+1}(x)$.

\item $B(x,d(x,y))\nsubseteq F_i(x)$.
\end{enumerate}

\item
Let $\mathcal{C}_{\textrm{ball}}(x,y,j)$ be the event that there
exists $i>0$ and $\mathcal{F}^{(i)} \in \mathcal{H}^{(i)}$ such that
the following holds:
\begin{enumerate}

\item $(\frac{8}{5})^j\le\rad(F_i(x))<(8/5)^{j+1}$.

\item $X_0 = B_{F_i(x)}(x_0,r)$ and $r$ chosen as in the algorithm.

\item $B(x,d(x,y))\bowtie(X_0,F_i(x)\setminus X_0)$.

\item $B(x,d(x,y))\subseteq F_{i+1}(x)$.

\end{enumerate}

Notice that by \claimref{claim:prob-diam-reduce} for each
$\mathcal{F}$, the first property holds for at most one value of
$i$, denote this value by $i_j(\mathcal{F})$.

\item
Let $\mathcal{X}(x,y,j,Z)$ be the event that there exist $i>0$ and
$\mathcal{F}^{(i)}\in \mathcal{H}^{(i)}$ such that the following
holds:
\begin{enumerate}
\item $(\frac{8}{5})^j\le\rad(F_i(x))<(8/5)^{j+1}$.

\item $Z = B_{F_i(x)}(x_0,r)$ and $r$ chosen as in the algorithm.

\item $B(x,d(x,y))\subseteq Z$.
\end{enumerate}

\item
Let $\mathcal{Y}(x,y,j,Z)$ be the event that there exist $i>0$
and $\mathcal{F}^{(i)}\in \mathcal{H}^{(i)}$  such that the
following holds:
\begin{enumerate}
\item $(\frac{8}{5})^j\le\rad(F_i(x))<(8/5)^{j+1}$.

\item $Z = F_i(x)\setminus B_{F_i(x)}(x_0,r)$ and $r$ chosen as in the algorithm.

\item $B(x,d(x,y))\subseteq Z$.
\end{enumerate}

\item
Let $\mathcal{Z}(x,y,j,Z)= \mathcal{Y}(x,y,j,Z) \cup
\mathcal{X}(x,y,j,Z)$.
\end{itemize}

We omit the parameters $x,y,j$ (or part of them) from $\mathcal{C},
\mathcal{C}_{\textrm{ball}}, \mathcal{X},\mathcal{Y},\mathcal{Z}$
when clear from context.
Here is an informal description of events $\mathcal{C},
\mathcal{C}_{\textrm{ball}}, \mathcal{X},\mathcal{Y},\mathcal{Z}$.
Fix $x,y$ and let $B=B(x,d(x,y))$. Event $\mathcal{C}(j)$ is the
event that the first time that $B$ is cut is when the parent cluster
has radius $\approx (8/5)^j$. Event $\mathcal{C}_{\textrm{ball}}(j)
$ is the event that the first time that $B$ is cut is by the central
ball given that that the parent cluster has radius $\approx
(8/5)^j$, observe that $\mathcal{C}_{\textrm{ball}}(j) \subseteq
\mathcal{C}(j)$. Event $\cup_Z \mathcal{Z}(j,Z)$ is the complement
of $\mathcal{C}(j)$. For each $Z$, event $\mathcal{Z}(j,Z)=
\mathcal{Y}(j,Z) \cup \mathcal{X}(j,Z)$; Event $\mathcal{X}(j,Z)$
(respectively $\mathcal{Y}(j,Z)$) is the event that $B$ is contained
inside (respectively, outside) the central ball of a cluster whose
radius is $\approx (8/5)^j$.

\begin{comment}
Let $\bar{\mathcal{D}}(x,y,j,Y)$ be the event that there exists
$Y_0\subseteq X$, $i>0$, $0<i'<i$ and $\mathcal{F}^{(i)} \in
\mathcal{H}^{(i)}$ such that the following holds:
\begin{enumerate}
\item $(\frac{8}{5})^j\le\rad(F_i(x))<(\frac{8}{5})^{j+1}$.

\item $F_{i'}(x)=Y$.

\item $Y_0 = F_{i'}(x) \setminus X_0$ where $X_0 = B(x_0,r)$ and $r$
chosen as in the algorithm.

\item $B(x,d(x,y))\subseteq Y_0$.
\end{enumerate}

\end{comment}

For each cluster we define the depth of its local density change as
a function of the ratio between its radius and its parent reset
radius. The parent reset cluster $Y_i(x)$ of a cluster $F_i(x)$ is
defined as follows. For any $i>0$ if $F_i(x)\in\mathcal{R}$ let
$Y_i(x)=F_i(x)$, otherwise let $Y_i(x)\in\mathcal{R}$ such that
$F_i(x)\in\mathcal{G}_{Y_i(x)}$. The depth of the local density
change is defined as
\begin{definition}
Let
$$
\alpha_i(x)=\frac{1}{f(\log(2\rad(Y_i(x))/\rad(F_i(x))))}
$$
notice that $\alpha$ is a uniform function over $\mathcal{F}$, \ie
if $u,v \in F_i(x)$ then $\alpha_i(u)=\alpha_i(v)$.
\end{definition}

Given this parameter we define the local density of a node $x$ in a
subgraph $Y_0$ as
\begin{definition}
Let $$
\rho_{Y_0}(x,i)=\frac{|Y_0|}{|B_{Y_0}(x,\alpha_i(x)\rad(F_i(x))/64)|}~.
$$
\end{definition}

We shall use the following Lemma from \cite{ABN06}
\begin{lemma}
\label{lem:dist-property} Let $(X,d)$ be a metric space and $Z
\subseteq X$. let $\chi \geq 2$ be a parameter. Given $0 < \Delta
< \diam(Z)$ and a center point $v \in Z$, there exists a
probability distribution over partitions $(S,\bS)$ of $Z$ such
that $S=B_{(Z,d)}(v,r)$, and $r$ is chosen from a probability
distribution in the interval $[\Delta/4,\Delta/2]$, such that for
any $\theta \in (0,1)$ satisfying $\theta \geq \chi^{-1}$, let
$\eta = \frac{1}{16} \ln (1/\theta)/\ln \chi$ then
% and $\bar{\theta} = \max \{ \theta,\chi^{-1} \}$.
 for any $x \in Z$, the following holds:
\begin{eqnarray*}
\lefteqn{\Pr[B_Z(x,\eta \Delta) \bowtie (S,\bS) ] \leq } \\
& &  (1-\theta) \left[\Pr[B_Z(x,\eta \Delta) \nsubseteq \bS] +
2\chi^{-2} \right] ..
\end{eqnarray*}
\end{lemma}

\begin{comment}

And prove the following variation of the Uniform Padding Lemma from
\cite{ABN06}.

\begin{lemma}
\label{lem:padded-star-partition} Let $(X,d)$ be a finite metric
space and $x_0\in X$. For any $r\in [1/2,5/8]\rad_{x_0}(X)$ let
$Y_0=X\setminus B(x_0,r)$. Let $\Delta\in(0,\rad(X)/4]$ and
$\hat{\delta}\in (0,1/e]$. There exists probabilistic partition of
$Y_0$ into cones, and a collection of uniform functions $\{ \xi_P:
X \rightarrow \{0,1\} | P \in {\cal P} \}$ and $\{ \hat{\eta}_P: X
\rightarrow \{0,1/\ln(1/\hat{\delta})\} | P \in {\cal P} \}$, such
that for any $\hat{\delta} \leq \delta \leq 1$, and
$\eta^{(\delta)}$ defined by $\eta_P^{(\delta)}(x) =
\hat{\eta}_P(x) \ln(1/\delta)$, the probabilistic partition
$\hat{\cal P}$ is $(\eta^{(\delta)},\delta)$-uniformly padded, and
the following conditions hold for any $P \in {\cal P}$ and any $x
\in X$:

\begin{itemize}

\item If $\xi_P(x)=1$ then: $2^{-6} /\ln
\rho(x,\bar{\Delta},\Gamma) \leq \hat{\eta}_{P}(x) \leq \\
2^{-6}/\ln(1/\hat{\delta})$.

\item If $\xi_P(x)=0$ then: $\hat{\eta}_{P}^{(\delta)}(x) = 2^{-6}
/\ln(1/\hat{\delta})$ and $\bar{\rho}(x,\bar{\Delta},\Gamma) \\
< 1/\hat{\delta}$.

\end{itemize}

\end{lemma}

\end{comment}

Given this lemma we prove a variant of the  Uniform Padding Lemma of
\cite{ABN06} that is tailored to the  construction of our algorithm.
There are three main differences. The first difference is that
instead of cutting balls we cut cones, the second difference is that
the parameter of the cut is defined in a subtle way with respect to
the last reset cluster: the local density change of a node is
defined as $\rho_{Y_0}(x,i)$ which depends on $\alpha_i(x)$ which
depends on $\rad(Y_i(x))/\rad(F_i(x))$ and \eqnref{eqn:f}. The final
difference is that the hierarchical scheme ensures the relation
$\frac{|F_i(x)|}{|Y_i(x)|} \le c\frac{\rad(F_i(x))}{\rad(Y_i(x))}$.

\begin{lemma}\label{lem:prob-cut}
For all $Y_0 \subset X$,  $x,y\in G_\epsilon$ and $j>0$ such that
$d(x,y)\le (\frac{8}{5})^j/(32f(\log(2c/\epsilon)))$:
\[
\Pr[\mathcal{C}(x,y,j)\mid\mathcal{Y}(x,y,j,Y_0)]
\le\frac{2^8d(x,y)\cdot f(\log(2c/\epsilon))}{(8/5)^j}\cdot
\ln\left(\frac{|Y_0|}
{|B_{Y_0}(x,(8/5)^j/(64f(\log(2c/\epsilon))))|}\right)~.
\]
\end{lemma}

\begin{proof}
Fix $Y_0 \subset X$,  $x,y\in G_\epsilon$ and $j>0$, such that
$d(x,y)\le (8/5)^j/(32f(\log(2c/\epsilon)))$. Let
$\mathcal{F}^{(i)}$ be any partial laminar family consistent with
the event $\mathcal{Y}(x,y,j,Y_0)$, hence
$i=i_j(\mathcal{F}^{(i)})$.

Now we bound the probability that $B(x,d(x,y)) \nsubseteq
F_{i+1}(x)$ given $\mathcal{F}^{(i)}$ and that the central ball
$X_0$ is disjoint from $B(x,d(x,y))$.

From $B(x,d(x,y))\subseteq F_i(x)$ follows $|F_i(x)|\ge\epsilon
n$. We know by the construction and definition of reset clusters
that $\frac{|F_i(x)|}{|Y_i(x)|}\le
c\frac{\rad(F_i(x))}{\rad(Y_i(x))}$ hence
$2\rad(Y_i)/\rad(F_i(x))\le 2c/\epsilon$ which implies that
$\alpha_i(x)\ge\frac{1}{f(\log(2c/\epsilon))}$.

Let $\Delta=\alpha_i(x)\rad(F_i(x))/4$ . For $k\ge 1$ let $v_k$,
$x_k$, $\hat{\chi}_k$ and $\chi_k$ be as in the algorithm, and let
$\ell_k$ be the appropriate cone-metric.

Let $\delta=\delta_{x,y,j,Y_0}=\exp\left\{-\frac{2^8d(x,y)\cdot
f(\log(2c/\epsilon))}{(8/5)^j}\cdot \ln\left(\frac{|Y_0|}
{|B_{Y_0}(x,(8/5)^j/(64f(\log(2c/\epsilon))))|}\right)\right\}$. If
$\delta < e^{-1}$ then the claim is trivial (probability is always
bounded by 1), so the interesting cases are when $\delta\ge e^{-1}$.
Let $\theta = \delta^{1/2}$. Note that $\theta \geq 2\chi_k^{-1}$ as
required (the algorithm actually apllies
\lemmaref{lem:dist-property} on $(Y_k,\ell_k)$ with $x_k$ as center
and the parameter $\chi_k$).

First consider the case that $\rho_{Y_0}(x,i)<2$, then we claim that
$B_{Y_0}(x,d(x,y))$ cannot be cut by a cone:

Since $v_1$ was chosen as to minimize $\rho_{Y_0}(z,i)$ then
$\rho_{Y_0}(v_1,i)<2$ as well. It implies that both\\
$|B_{Y_0}(v_1,\Delta/16)|,|B_{Y_0}(x,\Delta/16)|> |Y_0|/2$, hence
$B_{Y_0}(v_1,\Delta/16)\cap B_{Y_0}(x,\Delta/16)\neq\emptyset$,
therefore $d(x,v_1)\le \Delta/8$. Since $d(x,y)\le\Delta/8$ and
$\ell_1(v_1,x_1)=0$ follows that $B_{Y_0}(x,d(x,y))\subseteq
B_{(Y_0,\ell_1)}(x_1,\Delta/4)$.

Now assume that $\rho_{Y_0}(x,i)\ge 2$. We now claim that for all
$x\in Y_{k-1}$, $\eta_k\Delta\ge d(x,y)$. Recall that $\eta_k =
2^{-4}\ln(1/\theta)/\ln \chi_k=2^{-5}\ln(1/\delta)/\ln \chi_k$, and
notice that if $x\in Y_{k-1}$ then $\rho_{Y_0}(x,i)\ge\hat{\chi_k}$.
If $\hat{\chi}_k<4$ then $\chi_k=4$ and
$\log\rho_{Y_0}(x,i)/\log\chi_k\ge 1/2$, otherwise
$\chi_k=\hat{\chi}_k$ and $\log\rho_{Y_0}(x,i)/\log\chi_k\ge 1$.
Since $\alpha_i(x)\rad(F_i(x))\ge (8/5)^j/f(\log(2c/\epsilon))$ we
get:
\[
\eta_k\Delta\ge\frac{2^8d(x,y)f(\log(2c/\epsilon))\cdot\log\rho_{Y_0}(x,i)}{2^5(8/5)^j\log\chi_k}\cdot\frac{(8/5)^j}{4f(\log(2c/\epsilon))}
\ge d(x,y).
\]

It remains to show that if $x\in X_k$ then\\
$\Pr[B_{Y_0}(x,\eta_k\Delta)\nsubseteq X_k]\le 1-\delta\le
\frac{2^8d(x,y)\cdot f(\log(2c/\epsilon))}{(8/5)^j}\cdot
\ln\left(\frac{|Y_0|}
{|B_{Y_0}(x,(8/5)^j/(64f(\log(2c/\epsilon))))|}\right)$ as required.

Consider the distribution over partitions of $Y_0$ into cones $X_1,
X_2, \ldots X_t$ as defined above.  For $1 \leq m \leq t$, define
the events:
\begin{eqnarray*}
{\cal Z}_m & = & \{ \forall k, 1 \leq k < m, \,\, B_{Y_0}(x,\eta_k \Delta) \subseteq Y_k \}, \\
{\cal E}_m &  = & \{ \exists k, \, m \leq k < t \,\, {\rm s.t.} \,\,
B_{Y_0}(x,\eta_k \Delta) \bowtie (X_k,Y_k) | {\cal Z}_m \}.
\end{eqnarray*}
%Also let $T = B(x,2 \Delta)$.

We prove the following inductive claim: For every $1 \leq m \leq
t$:
\begin{eqnarray}
\label{eq:prob-event}
 \Pr[ {\cal E}_m ] \leq (1-\theta) (1 + \theta\E[ \sum_{k \geq m} \chi_k^{-1} |
 {\cal Z}_{m} ]).
\end{eqnarray}
The proof is essentially the same as the one in \cite{ABN06}.

Note that $\Pr[ {\cal E}_t ]=0$. Assume the claim holds for $m+1$
and we will prove for $m$. Define the events:
\begin{eqnarray*}
{\cal F}_m & = & \{ B_{Y_0}(x,\eta_m \Delta) \bowtie
(X_m,Y_m) | {\cal Z}_m \}, \\
{\cal G}_m & = & \{ B_{Y_0}(x,\eta_m \Delta) \subseteq Y_m | {\cal
Z}_m \} = \{ {\cal Z}_{m+1} | {\cal Z}_m \}.
\end{eqnarray*}
First we bound $\Pr[ {\cal F}_m ]$. Assume first a particular
choice of the cones $X_1, \ldots X_{m-1}$ such that event ${\cal
Z}_m$ occurs. Call this specific event ${\cal A}$, then given that
${\cal A}$ occurred the point $v_m$ is now determined
deterministically, and so is the value of $\chi_m$.
%Notice that
%since $r_m \leq \Delta$, if $B_Z(x,\eta_m \Delta) \bowtie
%(S_{v_m},\bS_{v_m})$ then $d(v_m,x) \leq \Delta + \eta_m \Delta
%\leq 2\Delta$, and thus $v_m \in T$.
Now, applying \lemmaref{lem:dist-property} we get
\begin{eqnarray*}
\lefteqn{
 \Pr[ B_{Y_0}(x,\eta_m \Delta) \bowtie
(X_m,Y_m) | {\cal A} ]  \leq } \\
& &  (1-\theta) (\Pr[B_{Y_0}(x,\eta_m \Delta) \nsubseteq Y_m | {\cal
A} ] + \theta\chi_m^{-1}.
\end{eqnarray*}
It follows that
\begin{eqnarray*}
\Pr[ {\cal F}_m ] \leq (1-\theta) (\Pr[ \bar{{\cal G}}_m ] +
 \theta \E[\chi_m^{-1} | {\cal Z}_m ]).
\end{eqnarray*}
Using the induction hypothesis we prove the inductive claim:
\begin{eqnarray*}
 \Pr[ {\cal E}_m ] & \leq & \Pr[ {\cal F}_m ] + \Pr[ {\cal G}_m ] \Pr[
 {\cal E}_{m+1} ] \\
 & \leq & (1-\theta) (\Pr[ \bar{{\cal G}}_m ] +
 \theta \E[\chi_m^{-1} | {\cal Z}_m ]) + \\
 & & \Pr[ {\cal G}_m ] \cdot (1-\theta) (1 + \theta \E[ \sum_{k \geq m+1} \chi_k^{-1} |
 {\cal Z}_{m+1} ]) \\
 & \leq & (1-\theta) (1 + \theta \E[ \sum_{k \geq m} \chi_k^{-1} |
 {\cal Z}_{m} ]),
\end{eqnarray*}
% proving the inductive claim.

Now consider a fixed choice of star-partition $\{X_0,\dots,X_t\}$.
Since the radius of every cone is at least $\Delta/4$, and since for
every $k\in [t]$, $\ell_k(v_k,x_k)=0$ we get that
$B_{(Y_0,d)}(v_k,\Delta/16)\subseteq
B_{(Y_0,\ell_k)}(x_k,\Delta/4)\subseteq X_k$. Therefore if $k\neq
k'$ then $B_{(Y_0,d)}(v_k,\Delta/16)\cap
B_{(Y_0,d)}(v_{k'},\Delta/16)=\emptyset$. Hence, we get:
\begin{eqnarray*}
\sum_{k \geq m} \chi_k^{-1} \leq \sum_{k \geq m} \hat{\chi_k}^{-1} =
\sum_{k \geq m} \frac{|B_{(Y_0,d)}(v_k,\Delta/16)|}{|Y_0|}\leq 1\,.
\end{eqnarray*}

We conclude that if $x\in X_m$
\begin{eqnarray*}
\lefteqn{ \Pr[B_{Y_0}(x,\eta_m\Delta) \nsubseteq X_m]
= \Pr[{\cal E}_1] \leq }\\
& & (1-\theta) (1 + \theta\cdot \E[\sum_{k \geq 1} \chi_k^{-1}]
)\leq (1-\theta)(1+\theta) = 1 -\delta.
\end{eqnarray*}

Since $\mathcal{Y}(x,y,j,Y_0)$ we have that $B(x,d(x,y))\subseteq
Y_0$, hence indeed $\Pr[B(x,d(x,y))\nsubseteq X_m]\le 1-\delta$.

\begin{comment}
It follows that $\hat{\cal P}$ is uniformly padded. Finally, we show
the properties states in the lemma. Recall that $\hat{\eta}_P(x) =
2^{-6}/\max\{\ln \hat{\chi}_j,\ln(1/\hat{\delta})\}$. By the choice
of $v_j$, $\hat{\chi}_j = \rho(Z_j,\bar{\Delta},\Gamma)$. As $x \in
Z_j$ we have that $\hat{\chi}_j \leq \rho(x,\bar{\Delta},\Gamma)$.
If follows that if $\xi_P(x)=1$ then $\hat{\chi_j} \geq
1/\hat{\delta}$ and therefore $\hat{\eta}_P(x) \geq 2^{-6}/\ln
\hat{\chi_j} \geq 2^{-6}\ln(1/\delta)/\ln
\rho(x,\bar{\Delta},\Gamma)$.

On the other hand if $\xi_P(x)=0$ then $\hat{\chi_j} <
1/\hat{\delta}$ and $\hat{\eta}_P(x)  = 2^{-6}/\ln(1/\hat{\delta})$.
Since $\diam(Z_j) \leq \diam(Z) \leq \bar{\Delta}$ we have that $Z_j
\subseteq B(x, \bar{\Delta})$, and therefore
$\bar{\rho}(x,\bar{\Delta},\Gamma) \leq \hat{\chi}_j <
1/\hat{\delta}$.

\end{comment}

\end{proof}

We complete the algorithm analysis be proving that the expected
distortion is scaling. As with many partition based schemes that use
local density, the \emph{core} argument is essentially based on the
observation that the series $\sum_{a<i \le b} \log
\frac{|B(x,2^i)|}{|B(x,2^{i-1})|}$ is a telescoping series hence it
can be bounded by $\log \frac{|B(x,2^b)|}{|B(x,2^{a})|}$. When
$|B(x,2^{a})| \geq \epsilon n$ and $b$ is large enough then this
argument gives the essential $O(\log 1/\epsilon)$ scaling
ingredient. The following is a technical generalization of this core
idea. The main problem is that the local density change
$\rho_{Y_0}(x,i)$ of our algorithm is defined as a function of
$Y_0$, but $Y_0$ is determined by a probabilistic processes. Hence
in order for the core telescoping argument to work we need to
delicately combine the various probabilistic events in a
hierarchical manner. This is done by induction.

\begin{lemma}
For any $x,y\in G_\epsilon$ we have
\[
\mathbb{E}[d_T(x,y)]\le\tilde{O}(\log^2(1/\epsilon))d(x,y).
\]
\end{lemma}

%D(j) - cut in level corresponding to radius j
%bar(D) - not cut in level corrsepond to eradus j.

\begin{proof}
For any $\epsilon>0$ fix some $x,y\in G_\epsilon$. Let $\ell$ be the
smallest integer such that $d(x,y)\le
(8/5)^\ell/(64f(\log(2c/\epsilon)))$, and let $\lceil
L=\log_{(8/5)}\diam(X)\rceil$. For ease of notation for any $j>0$
%
%and $Z_j\subseteq X$
writing $\E_{Z_j}$ means that the expectation is over clusters $Z_j$
such that $(\frac{8}{5})^j\le\rad(Z_j)<(8/5)^{j+1}$ that contain
$B(x,d(x,y))\subseteq Z_j$ whose distribution is induced by the
hierarchical probabilistic star partition algorithm.

% \alert{this is new, please check...(needed otherwise balls won't cancel out)}

%Define for any $j\ge\ell$ the event
%$\mathcal{D}(x,y,j)=\mathcal{D}(j)$ as $\exists m\in[\ell,\dots,j]$
%such that $\mathcal{C}(m)$.
Let $k=2^4c\cdot\log_{(8/5)}\log(1/\epsilon)$. First we prove by
induction on $j\ge\ell+k$ the following claim: For any $m\in
[\ell,j-1]$ let $h=\max\{m+1,j-k+1\}$, then for any $Z_j\subseteq
X$:
\begin{eqnarray}
\lefteqn{\sum_{m=\ell}^{j-1}\mathbb{E}_{Z_h}\left[\Pr\left[\mathcal{C}(m)\mid\mathcal{Z}(Z_h)\right]\mid
\mathcal{Z}(Z_j)\right]\cdot(8/5)^m} \label{eqn:induction}\\
& \le &  2^{10}c\cdot
d(x,y)f(\log(1/\epsilon))\sum_{i=j-k+1}^j\mathbb{E}_{Z_i}\left[\ln\left(\frac{|Z_i|}{\epsilon
n}\right)\mid\mathcal{Z}(Z_j)\right]. \nonumber
\end{eqnarray}
The base cases when $j=\ell+k$ is proved similarly to the induction
step and we leave it for the reader.

Assume the claim holds for $j$ and prove for $j+1$. Fix any
$Z_{j+1}\subseteq X$, for abbreviation let
$B_Z(x)=B_{Z}(x,(8/5)^j/(64f(\log(2c/\epsilon))))$. Let $p_j$ be the
probability that $B(x,d(x,y))\subseteq X_j$, where $X_j$ is the
central ball in the star partition of the cluster $Z_{j+1}$.
Consider first the last element in the summation:

\begin{eqnarray}
\lefteqn{\Pr\left[\mathcal{C}(j)\mid\mathcal{Z}(Z_{j+1})\right]\cdot(8/5)^j} \label{eqn:level-j}\\
&\le&\Pr\left[\mathcal{C}_{\textrm{ball}}(j)\mid\mathcal{Z}(Z_{j+1})\right](8/5)^j
+(1-p_j)\mathbb{E}_{Y_j}\left[\Pr[\mathcal{C}(j)\mid\mathcal{Y}(Y_j)](8/5)^j\mid
\mathcal{Z}(Z_{j+1})\right]..\nonumber
\end{eqnarray}

Consider the term
$\Pr\left[\mathcal{C}_{\textrm{ball}}(j)\mid\mathcal{Z}(Z_{j+1})\right]$.
We choose the radius $r$ of the central ball to be in the "sparsest"
of two disjoint strips around $x_0$: $(1/2,9/16)\rad(Z_{j+1})$ and
$(9/16,10/16)\rad(Z_{j+1})$, hence only one of them can contain more
than half of the points in $Z_{j+1}$, and we will choose $r$ from
the other one, which contains less than half of the points.

Moreover, the radius is actually in a sub-strip - \ie in the
interval $(1/2,1/2+1/32)\rad(Z_{j+1})$ or in
$(1/2+3/32,1/2+1/8)\rad(Z_{j+1})$. Hence if
$B(x,\alpha_{i_j}(x)\rad(Z_{j+1})/64)$ intersects one of these
sub-strips, it will be fully contained within the appropriate strip
(recall that $\alpha\le 1$), which suggest that if the
$B(x,\rad(Z_{j+1})/64)$ can be cut by the central ball, it contains
less than half of the points in $Z_{j+1}$, \ie
$\rho_{Z_{j+1}}(x,i_j)\ge 2$. We conclude that
\begin{eqnarray}
\lefteqn{\Pr\left[\mathcal{C}_{\textrm{ball}}(j)\mid\mathcal{Z}(Z_{j+1})\right](8/5)^j} \label{eqn:ball}\\
&\le& \frac{2d(x,y)}{\rad(Z_{j+1})/32}(8/5)^j\nonumber\\
&\le& 2^6d(x,y)\cdot\rho_{Z_{j+1}}(x,i_j) \nonumber\\
&\le& 2^7d(x,y)(p_j+(1-p_j))\cdot\ln\left(\frac{|Z_{j+1}|}
{|B_{Z_{j+1}}(x)|}\right)\nonumber\\
&\le& 2^9c\cdot d(x,y)f(\log(1/\epsilon)) \nonumber\\
&&\cdot\left(p_j\cdot\E_{X_j}\left[\ln\left(\frac{|Z_{j+1}|}{|B_{X_j}(x)|}\right)\mid\mathcal{Z}(Z_{j+1})\right]
+(1-p_j)\E_{Y_j}\left[\ln\left(\frac{|Z_{j+1}|}{|B_{Y_j}(x)|}\right)\mid\mathcal{Z}(Z_{j+1})\right]\right).\nonumber
\end{eqnarray}
In the third inequality we used that $\alpha_{i_j}(x)\ge
1/f(\log(2c/\epsilon))$, and in the last inequality we simply
reduced the size of $B_{Z_{j+1}}(x)$ and added expectations.

As for the term
$\mathbb{E}_{Y_j}\left[\Pr[\mathcal{C}(j)\mid\mathcal{Y}(Y_j)](8/5)^j\mid\mathcal{Z}(Z_{j+1})\right]$,
we apply Lemma~\ref{lem:prob-cut} which suggests that for any
$Y_j\subseteq Z_{j+1}$ it is bounded by $2^9c\cdot d(x,y)\cdot
f(\log(1/\epsilon))\cdot
\mathbb{E}_{Y_j}\left[\ln\left(\frac{|Z_{j+1}|}
{|B_{Y_j}(x)|}\right)\mid\mathcal{Z}(Z_{j+1})\right]$.

Now consider the reminder of the sum, let $h'=\max\{m+1,j-k+2\}$.
Since for any $m\in[\ell,j-1]$, $\E_{Z_h}[\cdot\mid
\mathcal{Z}(Z_{j+1})] = \E_{Z_j}(\E_{Z_h}[\cdot\mid
\mathcal{Z}(Z_{j})]\mid \mathcal{Z}(Z_{j+1}))$ we get that
\begin{eqnarray}
\lefteqn{\sum_{m=\ell}^{j-1}\mathbb{E}_{Z_{h'}}\left[\Pr\left[\mathcal{C}(m)\mid\mathcal{Z}(Z_{h'})\right]\mid
\mathcal{Z}(Z_{j+1})\right]\cdot(8/5)^m} \label{eqn:sum}\\
&=&\mathbb{E}_{Z_j}\left[\sum_{m=\ell}^{j-1}\mathbb{E}_{Z_h}\left[\Pr\left[\mathcal{C}(m)\mid\mathcal{Z}(Z_h)\right]\mid
\mathcal{Z}(Z_j)\right]\cdot(8/5)^m\mid\mathcal{Z}(Z_{j+1})\right]\nonumber\\
&=&p_j\cdot\mathbb{E}_{X_j}\left[\sum_{m=\ell}^{j-1}\mathbb{E}_{Z_h}\left[\Pr\left[\mathcal{C}(m)\mid\mathcal{Z}(Z_h)\right]\mid
\mathcal{X}(X_j)\right]\cdot(8/5)^m\mid\mathcal{Z}(Z_{j+1})\right]\nonumber\\
&&+(1-p_j)\cdot\mathbb{E}_{Y_j}\left[\sum_{m=\ell}^{j-1}\mathbb{E}_{Z_h}\left[\Pr\left[\mathcal{C}(m)\mid\mathcal{Z}(Z_h)\right]\mid
\mathcal{Y}(Y_j)\right]\cdot(8/5)^m\mid\mathcal{Z}(Z_{j+1})\right]\nonumber
\end{eqnarray}

Notice that $h'$ was changed to $h$, meaning that we added
expectation over level $j-k+1$ as well, this does not change the
value of the expression. Applying the induction hypothesis to
\eqnref{eqn:sum} yields
\begin{eqnarray}
\lefteqn{\sum_{m=\ell}^{j-1}\mathbb{E}_{Z_{h'}}\left[\Pr\left[\mathcal{C}(m)\mid\mathcal{Z}(Z_{h'})\right]\mid
\mathcal{Z}(Z_{j+1})\right]\cdot(8/5)^m} \label{eqn:sum-induction}\\
&\le& 2^{10}c\cdot
d(x,y)f(\log(1/\epsilon))p_j\cdot\mathbb{E}_{X_j}\left[\sum_{i=j-k+1}^j\mathbb{E}_{Z_i}\left[\ln\left(\frac{|Z_i|}{\epsilon
n}\right)\mid\mathcal{X}(X_j)\right]\mid\mathcal{Z}(Z_{j+1})\right]\nonumber\\
&&+2^{10}c\cdot d(x,y)f(\log(1/\epsilon))
(1-p_j)\mathbb{E}_{Y_j}\left[\sum_{i=j-k+1}^j\mathbb{E}_{Z_i}\left[\ln\left(\frac{|Z_i|}{\epsilon
n}\right)\mid\mathcal{Y}(Y_j)\right]\mid\mathcal{Z}\!(Z_{j+1})\right]\nonumber
\end{eqnarray}

We now have all the ingredients to prove the inductive claim of
\eqnref{eqn:induction}. For abbreviation let $W=2^9c\cdot
d(x,y)f(\log(1/\epsilon))$.

\begin{eqnarray*}
\lefteqn{\sum_{m=\ell}^j\mathbb{E}_{Z_{h'}}\left[\Pr\left[\mathcal{C}(m)\mid\mathcal{Z}(Z_{h'})\right]\mid
\mathcal{Z}(Z_{j+1})\right]\cdot(8/5)^m}\\
&\le&
W\cdot\left(p_j\cdot\E_{X_j}\left[\ln\left(\frac{|Z_{j+1}|}{|B_{X_j}(x)|}\right)\mid\mathcal{Z}(Z_{j+1})\right]
+(1-p_j)\E_{Y_j}\left[\ln\left(\frac{|Z_{j+1}|}{|B_{Y_j}(x)|}\right)\mid\mathcal{Z}(Z_{j+1})\right]\right)\\
&&+W\cdot
(1-p_j)\E_{Y_j}\left[\ln\left(\frac{|Z_{j+1}|}{|B_{Y_j}(x)|}\right)\mid\mathcal{Z}(Z_{j+1})\right]\\
&&+2W\cdot\left(p_j\cdot\!\mathbb{E}_{X_j}\!\left[\!\sum_{i=j-k+1}^j\!\mathbb{E}_{Z_i}\!\left[\!\ln\left(\frac{|Z_i|}{\epsilon
n}\right)\mid\mathcal{X}(X_j)\!\right]\!\mid\mathcal{Z}(Z_{j+1})\!\right]\right)\\
&&+2W\cdot\left(
(1\!-\!p_j)\!\mathbb{E}_{Y_j}\!\left[\!\sum_{i=j-k+1}^j\!\mathbb{E}_{Z_i}\!\left[\!\ln\left(\frac{|Z_i|}{\epsilon
n}\right)\!\mid\mathcal{Y}(Y_j)\!\right]\!\mid\mathcal{Z}\!(Z_{j+1})\!\right]\!\right)\\
& = &W\cdot p_j\cdot
\E_{X_j}\left[\ln\left(\frac{|Z_{j+1}|}{|B_{X_j}(x)|}\right)+\sum_{i=j-k+1}^j\!\mathbb{E}_{Z_i}\!\left[\!\ln\left(\frac{|Z_i|}{\epsilon
n}\right)\mid\mathcal{X}(X_j)\!\right]\mid\mathcal{Z}(Z_{j+1})\right]\\
&&+2W\cdot
(1-p_j)\E_{Y_j}\left[\ln\left(\frac{|Z_{j+1}|}{|B_{Y_j}(x)|}\right)+
\sum_{i=j-k+1}^j\!\mathbb{E}_{Z_i}\!\left[\!\ln\left(\frac{|Z_i|}{\epsilon
n}\right)\!\mid\mathcal{Y}(Y_j)\!\right]\mid\mathcal{Z}(Z_{j+1})\right]\\
&&+W\cdot
p_j\cdot\!\mathbb{E}_{X_j}\!\left[\!\sum_{i=j-k+1}^j\!\mathbb{E}_{Z_i}\!\left[\!\ln\left(\frac{|Z_i|}{\epsilon
n}\right)\mid\mathcal{X}(X_j)\!\right]\!\mid\mathcal{Z}(Z_{j+1})\!\right]\\
&\le&W\cdot
p_j\sum_{i=j-k+2}^{j+1}\!\mathbb{E}_{Z_i}\!\left[\!\ln\left(\frac{|Z_i|}{\epsilon
n}\right)\mid\mathcal{Z}(Z_{j+1})\right]+2W\cdot (1-p_j)
\sum_{i=j-k+2}^{j+1}\!\mathbb{E}_{Z_i}\!\left[\!\ln\left(\frac{|Z_i|}{\epsilon
n}\right)\mid\mathcal{Z}(Z_{j+1})\right]\\
&&+W\cdot
p_j\sum_{i=j-k+2}^{j+1}\mathbb{E}_{Z_i}\!\left[\!\ln\left(\frac{|Z_i|}{\epsilon
n}\right)\mid\mathcal{Z}(Z_{j+1})\!\right]\\
&\le& 2^{10}c\cdot d(x,y)f(\log(1/\epsilon))
\sum_{i=j-k+2}^{j+1}\mathbb{E}_{Z_i}\left[\ln\left(\frac{|Z_i|}{\epsilon
n}\right)\mid\mathcal{Z}(Z_{j+1})\right].
\end{eqnarray*}

The first inequality follows from \eqnref{eqn:level-j},
\eqnref{eqn:ball} and \eqnref{eqn:sum-induction}. The second
equality is just a re-ordering of terms. The third inequality is the
telescope argument, it holds since for any choice of $X_j\subseteq
Z_{j+1}$, and any choice of $Z_{j-k+1}\subseteq X_j$ by definition
$\rad(Z_{j-k+1})\le (5/8)^{k-1}\rad(X_j)
\le\frac{\rad(X_j)}{2^7f(\log(2c/\epsilon))}$, since $x\in
Z_{j-k+1}$ follows $Z_{j-k+1}\subseteq B_{X_j}(x)$. The argument for
$Y_j$ is similar. So the elements depending on $X_j$ and $Y_j$
cancel out, and we don't need the expectation on $X_j$ and $Y_j$
anymore.

Let $j=L+1$, $Z_j=X$, then applying \lemmaref{lem:prob-small-radius}
and \eqnref{eqn:induction} completes the proof.
\begin{eqnarray*}
\E[d_T(x,y)]&\le&\sum_{m=1}^L\Pr[\mathcal{C}_m]\cdot 2\rad(T[F_{i_m}(x)])\\
&\le& 4c'\sum_{m=1}^{\ell-1}(8/5)^m +
4c'\sum_{m=\ell}^L\E_{Z_h}\left[\Pr[\mathcal{C}_m\mid\mathcal{Z}(Z_h)]\mid\mathcal{Z}(X)\right](8/5)^m\\
&\le& 4c'(5/3)(8/5)^\ell + 2^{12}c\cdot c'\cdot
d(x,y)f(\log(1/\epsilon))
\sum_{i=L-k+1}^L\mathbb{E}_{Z_i}\left[\ln\left(\frac{|Z_i|}{\epsilon
n}\right)\right]\\
&\le& 8c'\cdot d(x,y)64f(\log(2c/\epsilon)) + 2^{12}c\cdot c'\cdot
d(x,y)f(\log(1/\epsilon))2^4c\cdot\log\log(1/\epsilon)
\ln(n/(\epsilon n))\\
&=&\tilde{O}\left(\log^2(1/\epsilon)\right)
\end{eqnarray*}

\end{proof}

\bibliographystyle{plain}

%%%%%%%%%%%%%%%%%%%%%%%%%%%%%%%%%%%%%%%%%%%%%%%%%%%%%%%%%%%%%%%%%%%%%%%%%%
\end{document}